\documentclass[aps,pre,twocolumn,showpacs,preprintnumbers,amsmath,amssymb]{revtex4}
\usepackage{graphicx}
\usepackage{graphicx,epstopdf,xcolor,soul,multirow}
\usepackage{dcolumn}
\usepackage{bm}
\usepackage{mathrsfs} 
\usepackage{amsmath} 
\usepackage[colorlinks=true,linkcolor=blue,citecolor=blue]{hyperref}%

\begin{document}

\title{Understanding the origin of extreme events in El Ni\~{n}o-Southern Oscillation}
\author{Arnob Ray$^1$}
\author{Sarbendu Rakshit $^1$}
\author{Gopal K. Basak $^2$}
\author{Syamal K. Dana$^{3,4}$}
\author{Dibakar Ghosh$^1$}\email{dibakar@isical.ac.in}
\affiliation{$^1$Physics and Applied Mathematics Unit, Indian Statistical Institute, Kolkata 700108, India\\
$^2$Stat-Math Unit, Indian Statistical Institute, Kolkata 700108, India\\
$^3$Department of Mathematics, Jadavpur University, Kolkata 700032, India\\
$^4$Division of Dynamics, Technical University of Lodz, 90-924 Lodz, Poland}

\date{\today}

\begin{abstract}
	We investigate a low-dimensional slow-fast model to understand the dynamical origin of   El Ni\~no-Southern Oscillation. A close inspection of the system dynamics   using several bifurcation plots reveals that a sudden large expansion of the attractor occurs at a critical system parameter via a type of interior crisis. This interior crisis  evolves through merging of  a cascade of period-doubling and period-adding  bifurcations that leads to the origin of occasional amplitude-modulated extremely large events. 
	More categorically, a situation similar to homoclinic chaos arises  near the critical point, however, atypical global instability evolves as a channel-like structure in phase space of the system that modulates  variability of amplitude and return time  of the occasional large events and makes a difference from the homoclinic chaos. The slow-fast timescale  of the low-dimensional model  plays an important role on the onset of  occasional extremely large events. Such extreme events are characterized by their heights when they exceed a threshold level measured by a mean-excess function. The probability density of events' height displays  multimodal distribution with an upper-bounded tail. We identify the dependence structure of interevent intervals  to understand the predictability of  return time of such extreme events using autoregressive integrated moving average model and  box plot analysis.
\end{abstract}

\maketitle

\section{Introduction}\label{intro}
\par  El Ni\~no-Southern Oscillation (ENSO) is one of the powerful climatic and oceanic events and it is a result of interannual climate variability due to an interaction between atmospheric and ocean circulations \cite{book35}. It has two phases: El Ni\~no, a warming phase of sea surface temperature (SST) in the eastern and central equatorial Pacific ocean and La Ni\~na, an occasional cooling of ocean surface waters in that area. El Ni\~no usually occurs once in the range of 2-7 years \cite{book4}. The key features of ENSO are its irregularity of occurrence and amplitude modulation due to ocean-atmospheric instability \cite{asm1,tim1}, which draw attention of the scientific community and planners over the last few decades for its catastrophic effect and socioeconomic impact. SST of eastern Pacific ocean is cold in the normal period along the west coast of South America, but sudden changes of wind patterns, ocean circulation patterns, and rise of the oceanic surface temperature of eastern Pacific are  observed during the El Ni\~no phase of ENSO episodes \cite{book7}. It becomes necessary to understand the extreme nature of ENSO from a purely dynamical system perspective.

\par Strong El Ni\~{n}o events have devastating effects on nature \cite{rain1}. During El Ni\~no periods, normal patterns of tropical precipitation are heavily disturbed and ENSO also incurs global climate change \cite{tna}. Drought may occur in Australia, Indonesia, India, Kenya, Morocco, Algeria, Canada, Mexico, etc., and floods in the central and eastern Pacific regions,  parts of South America close to Argentina, Chile, Peru, Ecuador, etc., during this phase. El Ni\~no significantly disrupts the local economy related to ecosystems, fisheries, and agriculture.

\par Enormous research efforts have  been exerted on developing an effective dynamical model dealing with the atmosphere-ocean interactions, exploring the decadal tropical climate variability \cite{tim1,jin0,tim2}. A mechanism of positive feedback and delayed negative feedback of the ocean-atmosphere interaction in the equatorial Pacific, illustrates the dynamic features of the ENSO \cite{fb11,fb1}.  A coupled atmosphere-ocean model has been studied to obtain self-sustained oscillations without anomalous external forces for understanding the  characteristic feature of the ENSO phenomenon. A low- dimensional model based on recharge mechanism was tried to explain the existence of decadal ENSO amplitude modulation and decadal changes in the tropical mean state \cite{tim2}. The low-order nonlinear ENSO model characterizes a strong asymmetry which is an intrinsic characteristic property of the ENSO phenomenon influenced by the nonlinear effect of  the subsurface temperature \cite{asm1}.

\par  A kind of irregular  spiking behavior  has been reported in the low-order ENSO dynamical model and more categorically, in the dynamical sense, a homoclinic connection to a saddle focus was found, which was responsible for this kind of complexity in dynamics \cite{tim2}. This explained the decadal recurrence of El Ni\~no  as chaotic mixed-mode oscillation (MMO) \cite{mmo1} switching irregularly between large- and small-amplitude oscillations. Although it was expected that predictability of MMO kind of  El Ni\~no and La Ni\~na situations could be possible, it was shown, using a modified low-dimensional slow-fast model that strong El Ni\~no events on decadal timescales were completely unpredictable \cite{unp}.  Still researchers are  in search of appropriate methods to address the question of predictability of El Ni\~no events to mitigate its harmful impact on the society \cite{ghil, ghil2,risk, ghil1, ghil3}.
\par 
In this paper, we investigate the low-dimensional slow-fast ENSO model proposed by Timmermann {\it et al.}\ \cite{tim2}
	to understand the origin of self-terminating occasional large spiking in the time evolution of the SST leading to a type of extreme events mainly from the
	dynamical system point of view. We have adopted numerical as well as analytical techniques to extend the earlier
	results \cite{tim2,mmo1} to develop further insights on the dynamical features of the system and explore possibility of extremely large events in the system with a parameter variation. We acknowledge the earlier observation \cite{tim2} that a homoclinic connection plays a crucial role in the origin of almost decadal recurrence of El Ni\~{n}o events of almost equal height, which we classify here, in our study, as chaotic MMOs. Additionally, we observe intermittent large spiking events in the temporal evolution of SST with variations in both return time and amplitude near a critical value of the bifurcation parameter. Such a kind of recurrent events, occurring rarely, is considered here as extreme events which has been focused here from the dynamical system perspective \cite{pisarchik,masolar1,cavalcante,lucarini}. Underlying mechanism of the extreme events, in our model, is a type of interior crisis \cite{fan,rajat2} that occurs via a collision of the period-doubling (PD) cascades and period-adding (PA) bifurcations at a critical system parameter and it is different from the typical interior crisis elaborated, in the literature \cite{cri1,cri2,leo,sorbendu}, as a collision of a chaotic orbit and a saddle orbit or saddle point.
 Experimental evidence of extreme rogue waves  has also been studied in nonlinear optics \cite{masolar1,masolar3}.  A closer inspection, in the region of this merging of PD cascade and PA bifurcation, reveals a global instability that exists in the form of a channel-like structure \cite{rajat2,masolar2,masolar4} in state space of the system. The volume of this channel introduces a variability of amplitude and  return time of  intermittent large spiking events, which we classify here as extreme events and are clearly different from the El Ni\~no events, which information is missing  in the earlier  report \cite{tim2, mmo1}. We present numerical evidence that the slow-fast timescale of the model system plays a crucial role in the onset of such extreme events in the model. Extreme events are identified here using a mean excess function  \cite{cole} and thereby, we also estimate the probability density function (PDF) of events' height as usually done for extreme events, in general, in single or coupled systems \cite{masolar1,leo,de,mishra, arnob}.  We confirm rare occurrence of  this new kind of extreme event (large variability of height and return time) using their  interval of return time. We have collected interevent interval data from numerical experiments and plotted histograms of return time and then try to fit with the known distributions, Weibull, Gamma, and Log-normal distributions for a comparison and thereby to find an appropriate distribution of events. Besides, we try to address the question of predictability of extreme events using autoregressive integrated moving average (ARIMA) model for time-series forecasting \cite{chatfield} and box plot analysis \cite{box} of interevent interval of extreme events.
\par This paper is organized as follows. In Sec.\ \ref{model}, we briefly describe the slow-fast ENSO model. In Sec.\ \ref{dynamics}, we explain the system dynamics and try to understand the mechanism of extreme events in Sec.\ \ref{origin}. In Sec.\ \ref{stat}, we characterize the extreme events elaborately using statistical tools and address the question of predictability of the observed extreme events in Sec.\ \ref{predict}.  Finally, discuss an important point in Sec.\ \ref{timescale} that the slow-fast parameter plays a crucial role on the onset of extreme events. Results are summarized in Sec.\ \ref{conclu}.


\section{Slow-fast ENSO Model}\label{model}
\par A general formulation of a slow-fast dynamical system \cite{jg} can be written as
\begin{equation}
\begin{array}{lll}
\label{eq.1}				
\eta \dfrac{dx}{dt}=f(x,y,\lambda); 
~\dfrac{dy}{dt}= g(x,y,\lambda),
\end{array}
\end{equation}
where $x\in \mathbb{R}^{m} \;\mbox{and}\; y\in \mathbb{R}^{n}$ are fast and slow variables, respectively, $\lambda\in \mathbb{R}^{p}$ is a model parameter, and $0<\eta<<1$ represents the ratio of timescales.	Setting $\eta$=0, the trajectory of Eq.\ (\ref{eq.1}) converges to the solution of differential algebraic equation $f(x,y,\lambda) =0$ and $\dfrac {dy}{dt}=g(x,y,\lambda)$, where $\mathcal{S}= \{(x,y) \in \mathbb{R}^{m} \times \mathbb{R}^{n}| f(x,y,\lambda) =0\}$ is a critical manifold. Now we introduce the slow-fast ENSO model proposed by Timmermann {\it et al.}\ \cite{tim2}. If $h_{1}$ is the thermocline depth of the western Pacific, then its evolution equation is,
\begin{equation}
\begin{array}{l}\label{eq.2}	
\dfrac{dh_1}{dt}=r\bigg(-h_1-\dfrac{bL w_s}{2}\bigg),
\end{array}
\end{equation}
where $r$ represents the dynamical adjustment timescale, $w_s$ denotes the zonal wind stress, $b$ is the efficiency of $w_s$ in driving thermocline slope, and $L$ is the basin width. The thermocline depth $h_{2}$ of the eastern Pacific is related to $h_{1}$ by the relation $h_{2}=h_{1}+bL w_s$.

\par The dynamical equations of equatorial sea surface temperatures $T_{1}$ and $T_{2}$ of the western and eastern Pacific, respectively, are represented by
\begin{equation}
\begin{array}{l}\label{eq.3}
\dfrac{dT_1}{dt}=-\alpha(T_{1}-T_{r})-\dfrac{u(T_{2}-T_{1})}{\frac{L}{2}},\\[10pt]
\dfrac{dT_2}{dt}=-\alpha(T_{2}-T_{r})-\dfrac{w\big(T_{2}-T_{sub}(h_{1},T_{1},T_{2})\big)}{H_{m}},
\end{array}
\end{equation}
where $\frac{1}{\alpha}$ measures a typical thermal damping timescale, $T_{r}$ is the thermal relaxation towards a radiative-convective equilibrium temperature, $H_m$ denotes depth of the mixed layer, $u$ and $w$ are the zonal advection velocity and equatorial upwelling velocity, respectively. $T_{sub}$ is the subsurface temperature being upwelled into the mixed layer.

\par The zonal wind stress is related to SST as $w_s=-\frac{\mu(T_{1}-T_{2})}{\beta}$, where $\beta$ and $\mu$ are the coupling coefficient between SST and wind stress. Hence the expression of $h_2$ becomes
\begin{equation}
\begin{array}{l}\label{eq.4}
h_{2}=h_{1}+\dfrac{bL\mu(T_{2}-T_{1})}{\beta}.
\end{array}
\end{equation}
The zonal advection velocity $u$ and equatorial upwelling velocity $w$ are assumed to be proportional to the zonal wind stress anomalies $w_s$ as
\begin{equation}
\begin{array}{l}\label{eq.5}
~~\dfrac{u}{\frac{L}{2}}=\epsilon\beta w_s; \quad
\dfrac{w}{H_{m}}= -\zeta\beta w_s.
\end{array}
\end{equation}
Here $\epsilon$ and $\zeta$, respectively, quantify the strength of zonal and vertical advection. The expression of $T_{sub}$ can be written as
\begin{widetext}
	\begin{equation}
	\begin{array}{l}\label{eq.6}
	T_{sub}(h_1, T_1, T_2)= T_{r}-\dfrac{1}{2}(T_{r}-T_{r0})\Bigg[1-\tanh\dfrac{\Big(H+h_{1}+\frac{bL\mu(T_{2}-T_{1})}{\beta}-z_{0}\Big)}{h_*}\Bigg].
	\end{array}
	\end{equation}
\end{widetext}
Here $T_{r0}$ and $H$, respectively, are the mean eastern equatorial temperature and eastern thermocline reference depth, $z_{0}$ measures the depth at which  upwilling velocity $w$ takes its characteristic value, and $h_{*}$ is sharpness of thermocline. Hence we obtain the three-dimensional (3D) slow-fast  model,
\begin{widetext}
	\begin{subequations}\label{eq.7}
		\begin{align}
		\label{eq:7a}
		\frac{dh_1}{dt}&=r\bigg(-h_{1}-\dfrac{bL\mu(T_{2}-T_{1})}{2\beta}\bigg),\\
		\label{eq:7b}
		\frac{dT_1}{dt}&=-\alpha(T_{1}-T_{r})-\epsilon\mu(T_{2}-T_{1})^2,\\
		\label{eq:7c}
		\frac{dT_2}{dt}&=-\alpha(T_{2}-T_{r})+\zeta\mu(T_{2}-T_{1})\Bigg(T_{2}-T_{r}+\dfrac{1}{2}(T_{r}-T_{r0})\bigg[1-\tanh\frac{\Big(H+h_{1}+\frac{bL\mu(T_{2}-T_{1})}{\beta}-z_{0}\Big)}{h_*}\bigg]\Bigg),
		\end{align}
	\end{subequations}
\end{widetext}
when  $h_{1}$ represents a slow variable for $r<1$ and, $T_{1}$ and $T_{2}$ are fast variables. Throughout this paper, we make a choice of fixed parameters \cite{tim2},  $\zeta$=1.3, $r=\frac{1}{400}$~day$^{-1}$, $\alpha=1/180$~day$^{-1}$, $T_{r}=29.5^{\circ}$C, $L=15\times10^6$ m, $H_m=50$ m, $\dfrac{bL\mu}{\beta}=22mK^{-1}$, $\mu=0.0026K^{-1}\text{day}^{-1}$, $H=100$ m, $h_{*}=62$ m, $T_{r0}=16^{\circ}$ C, and $z_{0}=75$ m. We analyze the linear stability of the equilibrium points of system (\ref{eq.7}) and calculate the invariant manifold as detailed in the Appendix A. The system has one saddle $(0,T_{r},T_{r})$ and an interior equilibrium point $(h_{1}^*,T_{1}^*,T_{2}^*)$, which is a saddle focus. We integrate the system (\ref{eq.7}) using fourth-order Runge-Kutta algorithm and fixed time step $0.01$. Our main emphasis is to investigate the complex dynamical behavior of the system (\ref{eq.7}) by varying the strength of zonal advection $\epsilon$.
\begin{figure*}[ht]
	\centerline{\includegraphics[scale=0.3]{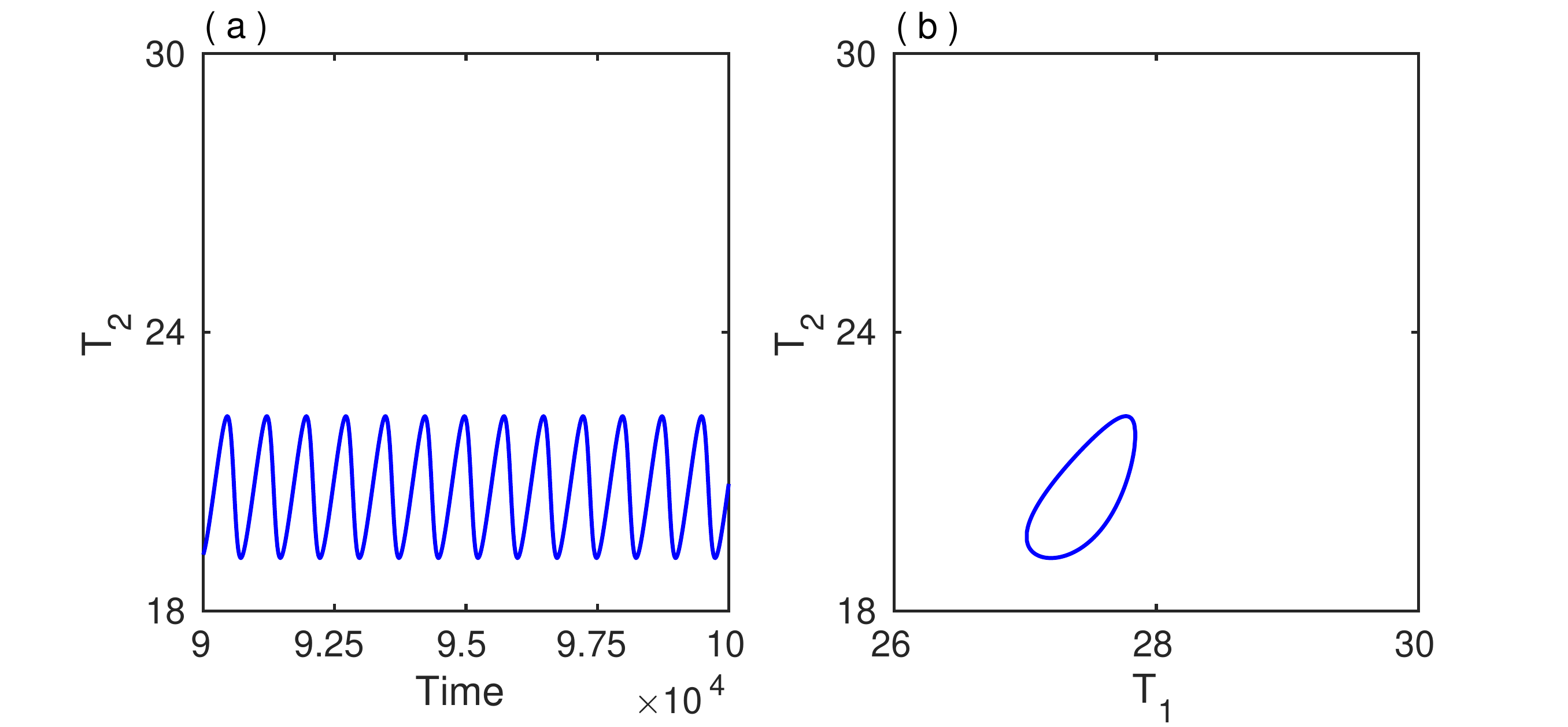}		\includegraphics[scale=0.3]{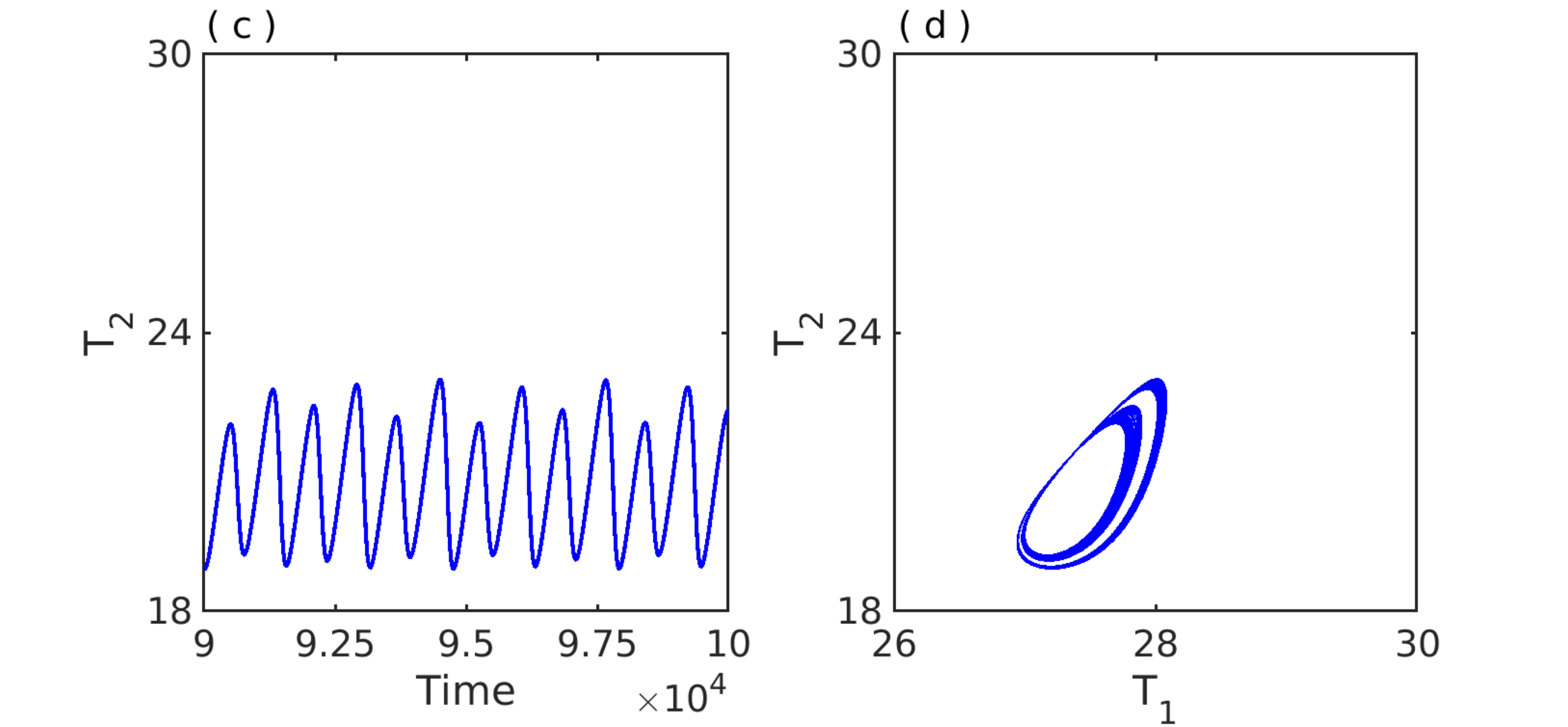}}
	\centerline{\includegraphics[scale=0.3]{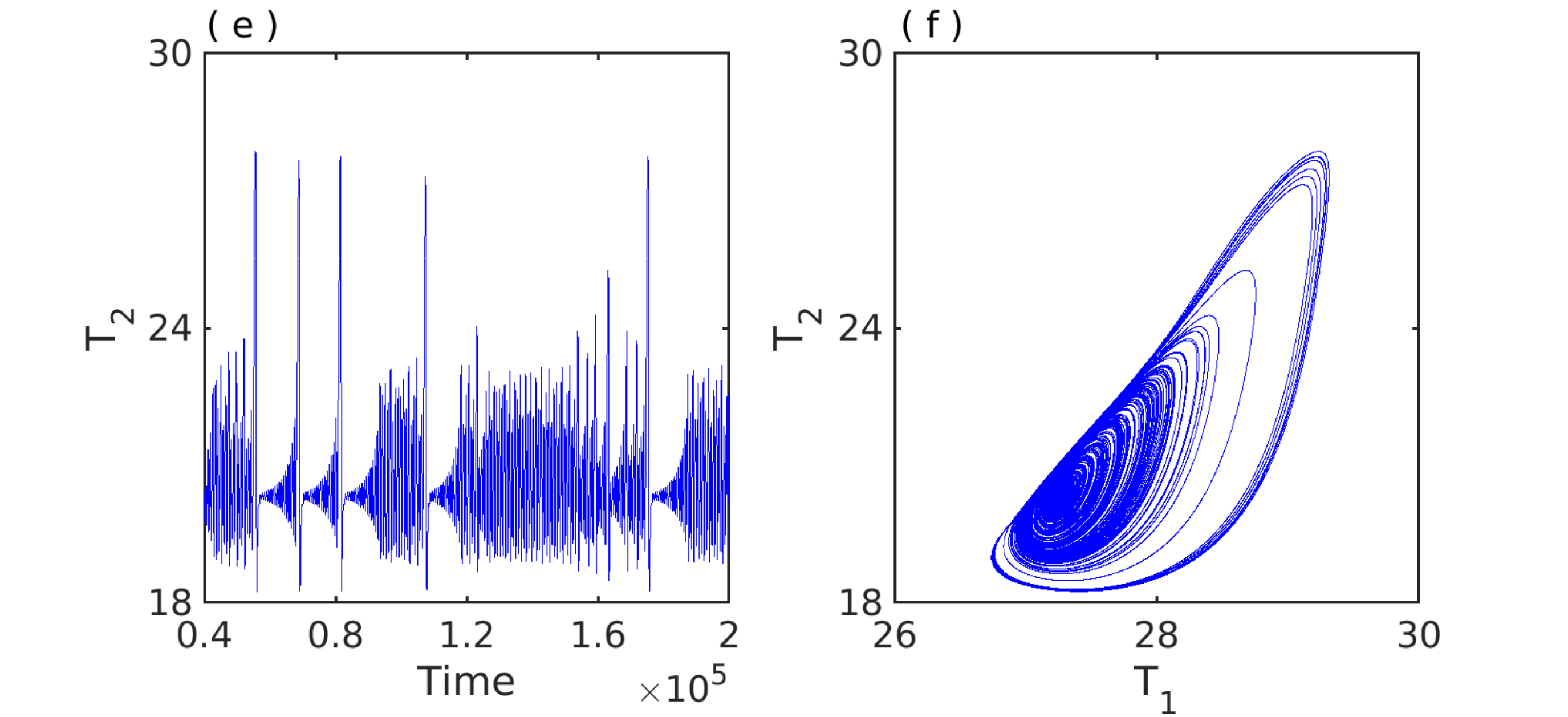}		\includegraphics[scale=0.3]{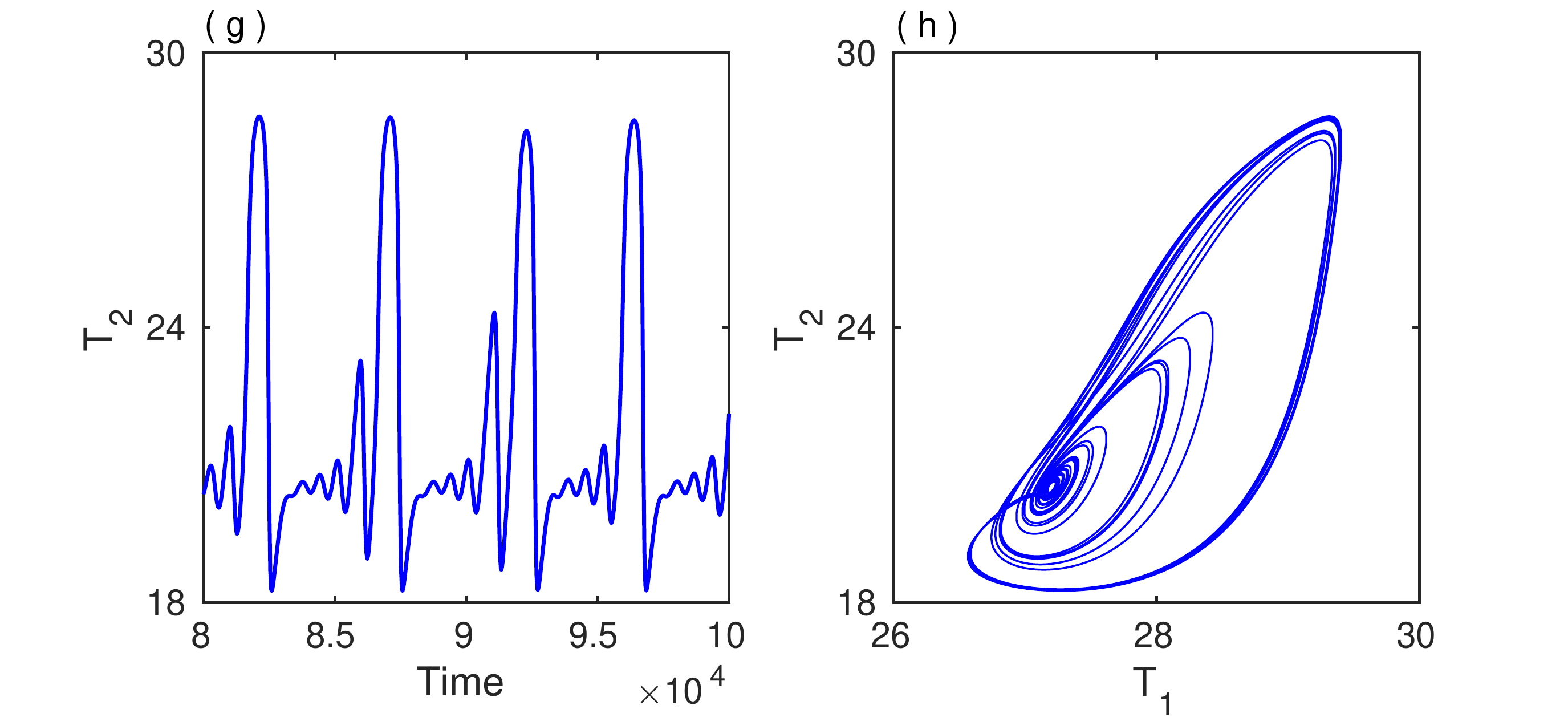}}
	\centerline{\includegraphics[scale=0.3]{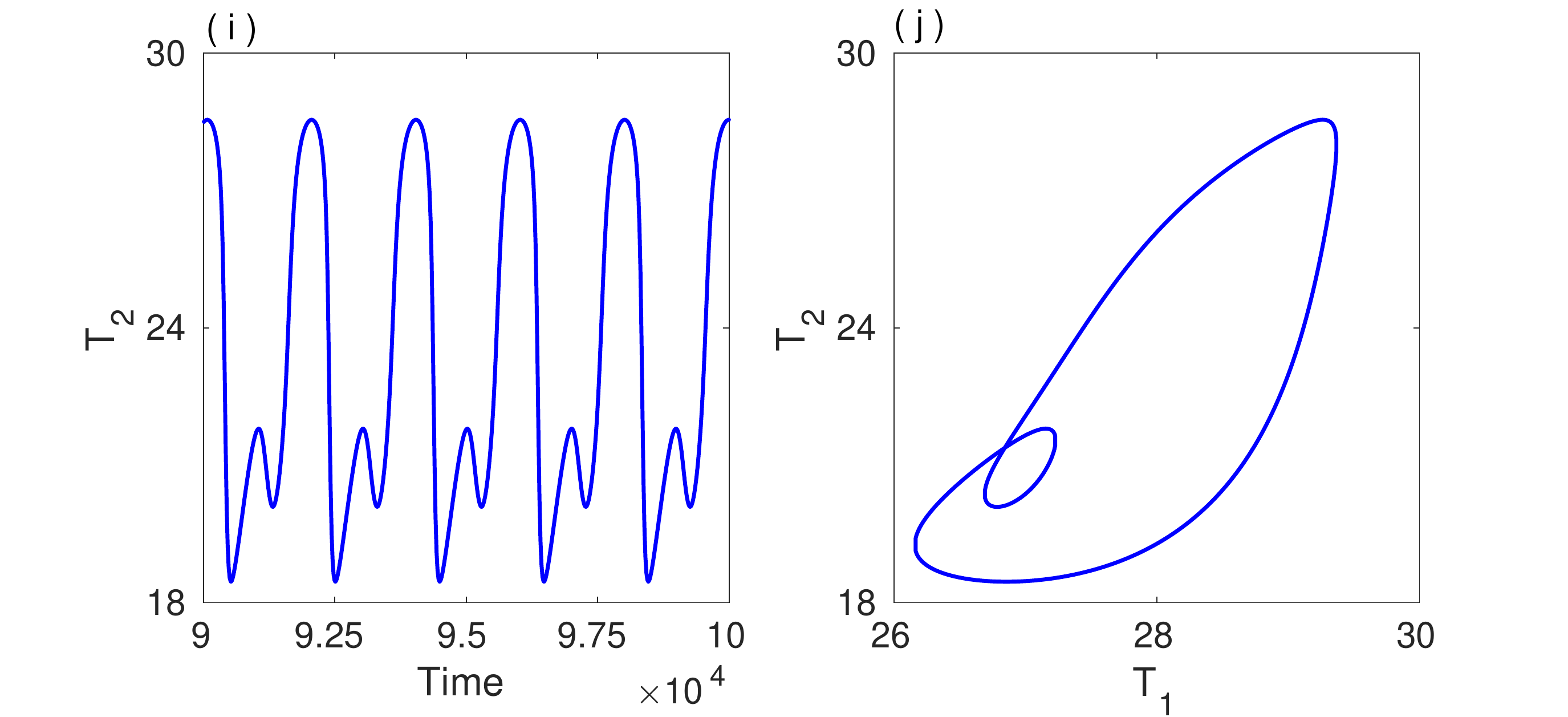}		\includegraphics[scale=0.3]{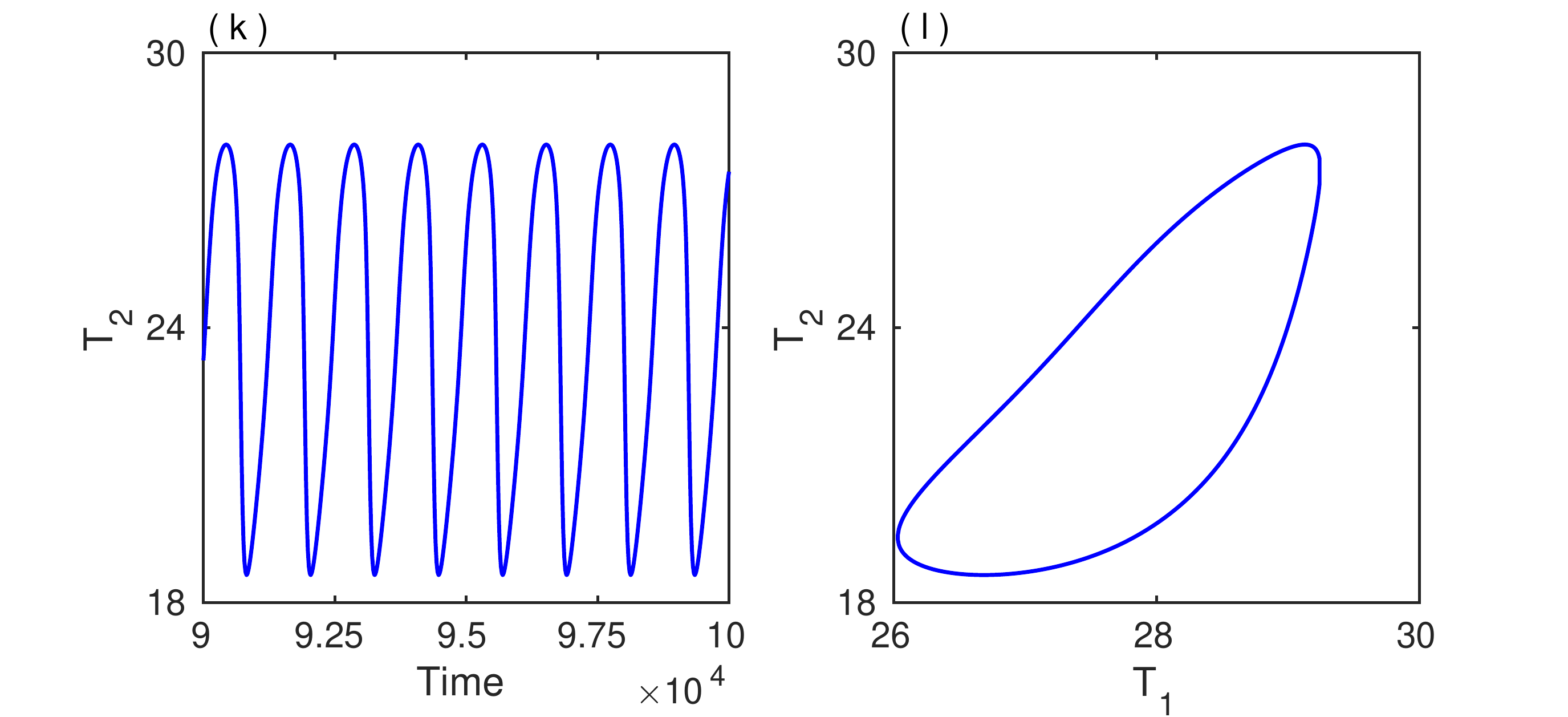}}
	\caption{Evolution of different dynamics in ENSO model with a variation of $\epsilon$. Temporal evolution of $T_2$ and corresponding phase space diagram in $(T_1,T_2)$ plane: [(a) and (b)]~ small-amplitude period-1 limit cycle at $\epsilon=0.0961$, [(c) and (d)]~bounded chaos at $\epsilon=0.0984$, [(e) and (f)]~ extreme events at $\epsilon=0.0985$, [(g) and (h)]~ chaotic MMO (or El Ni\~no events) at $\epsilon=0.1076$, [(i) and (j)]~ periodic MMO ($1^1$) at $\epsilon=0.15$, and [(k) and (l)]~ large-amplitude period-1 limit cycle at $\epsilon=0.17$.}
	\label{figure_1}
\end{figure*}
\section{Complex dynamics of ENSO model}\label{dynamics}

\par Emergence of El Ni\~no and extreme events is shown in Fig.\ \ref{figure_1} with a series of temporal dynamics and their phase portraits in a range of $\epsilon\in[0.0961,0.17]$. Figures\ \ref{figure_1}(a) and \ref{figure_1}(b) show a period-1 oscillation at $\epsilon=0.0961$, which evolves into  bounded chaos shown in Figs.\ \ref{figure_1}(c) and \ref{figure_1}(d) for $\epsilon=0.0984$ via a cascade of PD bifurcations (cf. bifurcation diagrams in Sec.\ \ref{origin} and in the Appendix A). It is noticeable that the amplitude of $T_2$ remains bounded ($T_{2_{max}}<24$) here, however, for a small increase beyond $\epsilon\approx0.09848$,  intermittent large spiking events ($T_{2_{max}}>24$) start appearing as shown in Fig.\ \ref{figure_1}(e) for $\epsilon=0.0985$, which occasionally visit a close vicinity of the interior equilibrium point, namely a saddle focus. This sudden change in amplitude of the oscillation and  occasional switching to small-amplitude oscillation is reflected  in the comparative phase portraits in  Figs.\ \ref{figure_1}(d) and \ref{figure_1}(f). The system trajectory in  Fig.\ \ref{figure_1}(f) spirals out in a 2D unstable manifold  of the saddle focus. The trajectory moves slowly near the saddle focus and it is trapped there for a while before finally spiraling  away from it. After moving away from the saddle focus, the trajectory attempts a global excursion to originate a large spike; however, it is reinjected  along the stable eigendirection to reach another close vicinity of the saddle focus with a highly irregular interval of time. The  time spent during spiraling out varies and it depends on how close the reinjected trajectory reaches a vicinity of the saddle focus and thereby it makes a large variation in the number of small oscillations that makes irregular return time of the large events. A tendency to develop a homoclinic chaos with a local instability of the saddle focus is seen here, but the typical global stability of homoclinic chaos \cite{dana} is never achieved due to the presence of a channel like structure. 
 The trajectory revolves around the saddle focus for quite some time due to local instability (attracted along the stable eigendirection, pushed away along the unstable eigenplane of the saddle focus) while a globally instability due to the channel-like structure induces a variation of the return path (global  excursion) of the trajectory making a wide variation in both the amplitude and return time of large spikes. 
Our main focus is the origin of this exceptional case of large variation in return time and amplitude of extreme events  illustrated in Figs.\ \ref{figure_1}(e) and \ref{figure_1}(f), which has not been reported so far, to the best of our knowledge. For higher values of $\epsilon=0.1076$ in Figs.\ \ref{figure_1}(g) and \ref{figure_1}(h), the dynamics becomes almost regular, but in reality, the oscillation has  low variability in amplitude of large spikes and the number of small oscillations still varies irregularly confirming existence of local instability, a typical signature of chaotic MMO reported earlier \cite{tim2,mmo1} as El Ni\~no events. In that particular case, large events are more frequent compared to our observed new kind of   extreme events shown in Fig.\ \ref{figure_1}(e). Following a series  of dynamical events with varying $\epsilon$ we notice a transition from chaotic MMOs to periodic MMOs ($1^1$)  shown in Figs.\ \ref{figure_1}(i) and \ref{figure_1}(j) for a larger $\epsilon=0.15$. The period-1 limit cycle returns for larger $\epsilon$, but with a larger amplitude of oscillation as shown in Figs.\ \ref{figure_1}(k) and \ref{figure_1}(l).
\begin{figure*}[ht]
	\centerline {\includegraphics[scale=0.5]{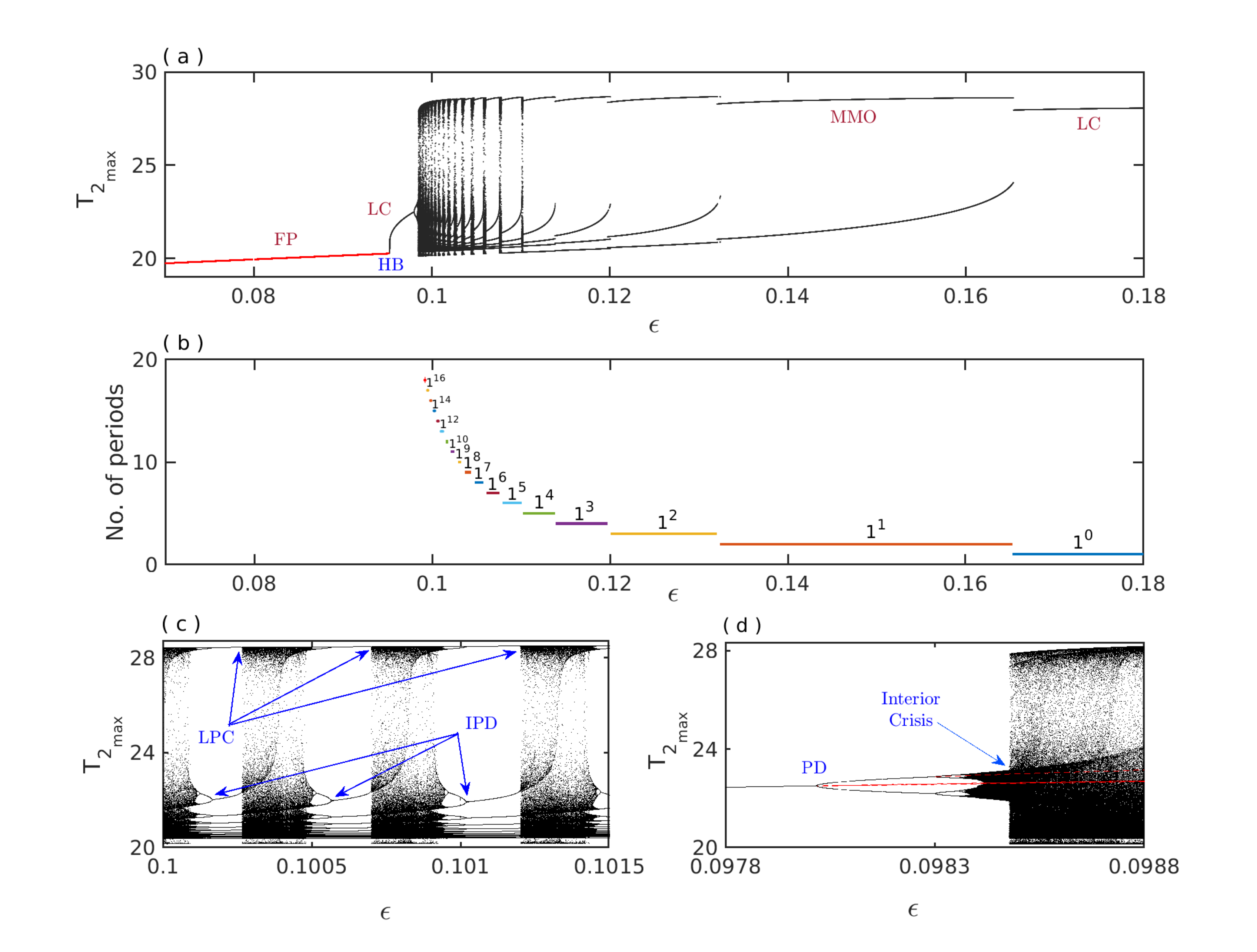}}
	\caption{(a) Bifurcation diagram of maximum peak values of $T_2$ against $\epsilon$. It shows different kind of oscillatory behavior of the system. (b) Period-parameter bifurcation: Period of periodic mixed mode oscillations are plotted with intervals. Zoomed version of the bifurcation diagram (c) for $\epsilon\in[0.1,0.1015]$ to get clear view of the inverse period-doubling (IPD) bifurcation and saddle-node bifurcation of limit point cycles (LPC)  and (d) for $\epsilon\in[0.0978,0.0988]$ to emphasize the interior crisis of the system.}
	\label{figure_2}
\end{figure*}
\section{Origin of extreme events}\label{origin}
\par For a closer inspection of the dynamical  evolution of extreme events with both local and global instabilities in the system, a bifurcation diagram of  $T_{2_{max}}$ against $\epsilon$ is drawn in Fig.~\ref{figure_2}(a).  The stable interior equilibrium (FP) evolves into a limit cycle (LC) via Hopf bifurcation (HB) and it continues until $\epsilon\approx0.09802$. The system undergoes a cascade of PD  bifurcations to become chaotic with  increasing $\epsilon$ (a broader scenario in parameter space presented in Fig.~\ref{figure_9} in  Appendix A).  At a critical value $\epsilon\approx0.09848$, $T_{2_{max}}$ suddenly increases to a large value where the cascading sequence of PD merges with  a PA sequence. The bounded chaotic mode explodes here when we notice intermittent large-amplitude spiking oscillations with a variation in amplitude intercepted by a varying number of small oscillations as shown in the temporal evolution of $T_2$ in Fig.\ \ref{figure_1}(e).  We classify such intermittent large events as the extreme events, which are followed by the typical chaotic MMOs identified as El Ni\~no events \cite{mmo1} and MMOs as shown in Fig.\ \ref{figure_2}(c)  [a zoomed version of Fig.\ \ref{figure_2}(a)] for increasing $\epsilon$; exemplary temporal  evolution of chaotic MMO and periodic MMO are shown in Figs.\ \ref{figure_1}(g) and \ref{figure_1}(i), respectively. For larger $\epsilon$, the dynamics return to LC, but with a large-amplitude as shown in Fig.\ \ref{figure_1}(k). The periodic MMO sequence is better visualized from the right side of the bifurcation diagram in Fig.\ \ref{figure_2}(a) when we decrease $\epsilon$ and observe a Farey sequence $L^s$ \cite{dana}  consisting of large-amplitude oscillations (denoted here by $L=1$) followed by small-amplitude oscillations (denoted by $s=1,2,3,..., \infty$). The Farey sequence of periodic MMOs emerges in successive parameter windows intermediate to chaotic windows of the control parameter $\epsilon$. Figure\ \ref{figure_2}(b) shows the sequence of MMOs ($1^1-1^2-...-1^{16}$) in a period-parameter bifurcation plot. It first emerges with a period-1 limit cycle ($1^0$) from $\epsilon\approx0.1653$ and continues until we see a $1^{16}$ MMO that shows a devil's staircase \cite{dana} with an asymptotic increase in the  time period of oscillations. From our numerical simulations, we are able to record  a MMO with a maximum number of $s=16$ (with our best effort) near $\epsilon=0.1$, although MMOs with larger $s$ values really exist in the system. The parameter window  of $\epsilon$ for each MMO becomes narrower with increasing $s$ and the time period of MMOs increases asymptotically. The interval between any two successive MMO windows in $\epsilon$ [Fig.~\ref{figure_2}(b)] is a chaotic window and hence the MMOs emerge in a sequence of alternate PA bifurcations. As we decrease $\epsilon$, we find an increasing number of small oscillations $s=1,2,3,...,16$ in the periodic MMO windows until the sequence collides with bounded chaos.

\par To make a clear picture of this scenario, a small range of $\epsilon$ values is zoomed in two separate bifurcation diagrams in Figs.~\ref{figure_2}(c) and \ref{figure_2}(d). Figure\ \ref{figure_2}(c) shows an example of PA sequence  in the range of  $\epsilon\in[0.1,0.1015]$. MMOs lie in the periodic windows where the chaotic state becomes periodic via inverse period-doubling (IPD) cascades  and  it becomes chaotic via saddle-node bifurcation. MMOs continue with alternate chaotic and periodic windows with decreasing $\epsilon$ until it reaches a chaotic MMO ($1^{\infty}$) and arrive at a critical point of transition where the amplitude drops down to small-amplitude bounded chaos at $\epsilon\approx0.09848$ as shown in Fig.\ \ref{figure_2}(d). Another view of the bifurcation scenario is presented in Fig.\ \ref{figure_2}(d) that is a zoomed version of Fig.~\ref{figure_2}(a) in the range of $\epsilon\in[0.0978, 0.0988]$ that focuses on the PD cascades to bounded chaos and a sudden large rise (fall) of amplitude of the chaotic oscillation as viewed for an increasing (decreasing) $\epsilon$.  This sudden change in amplitude of chaos occurs near $\epsilon\approx0.09848$ via the interior crisis due to a merging of PD and a PA bifurcation sequences \cite{rajat2,fan} and, it indicates onset of extreme events as shown in Fig.\ \ref{figure_1}(e). We observe the typical El Ni\~no events as chaotic MMOs for a larger $\epsilon$ value as shown in Fig.~\ref{figure_1}(g) and as reported earlier \cite{mmo1}.  For  a macroscopic view, we draw the critical manifold of the slow-fast system and search for the regions of instabilities in 3D phase space that provides a minefield  for the generation of extreme events. The critical manifold is obtained by solving a truncated system of the fast variables $T_1$ and $T_2$, given by the Eqs.\ (\ref{eq.7}b) and (\ref{eq.7}c),
\begin{widetext}
	\begin{equation}
	\begin{array}{l}\label{eq.8}
	\mathcal{S}=\Bigg\{(h_1,T_1,T_2)\in\mathbb{R}^3~\bigg|~\alpha-\mu \Bigg(\epsilon(T_{2}-T_{1})+\zeta\Big(T_{2}-T_{r}+\dfrac{1}{2}(T_{r}-T_{r0})\Big[1-\tanh\frac{H+h_{1}+\frac{bL\mu(T_{2}-T_{1})}{\beta}-z_{0}}{h_{*}}\Big]\Big)\Bigg)=0\Bigg\}.
	\end{array}
	\end{equation}
\end{widetext}

\begin{figure*}[ht]
	\centerline{\includegraphics[scale=0.25]{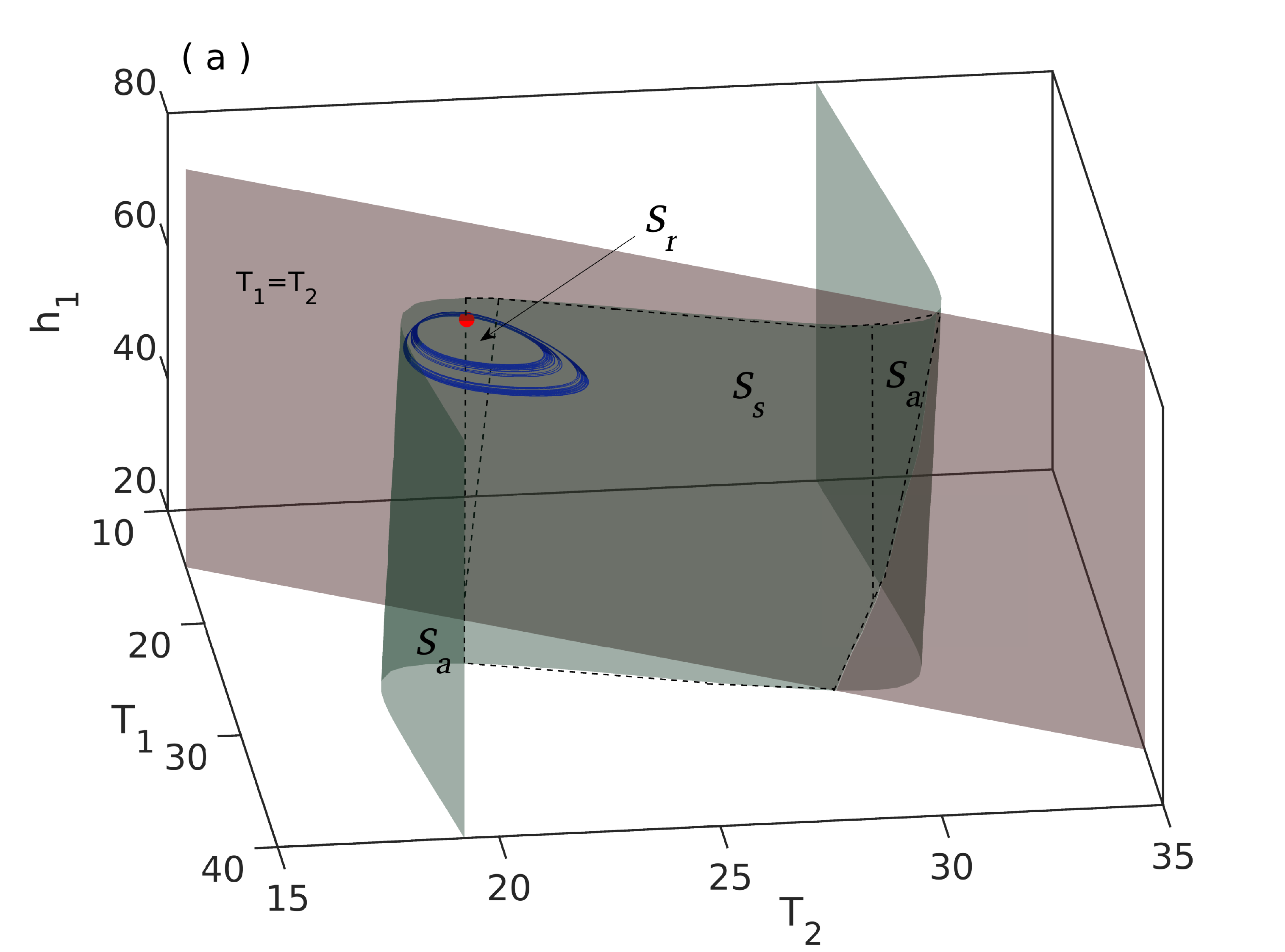}
		\includegraphics[scale=0.25]{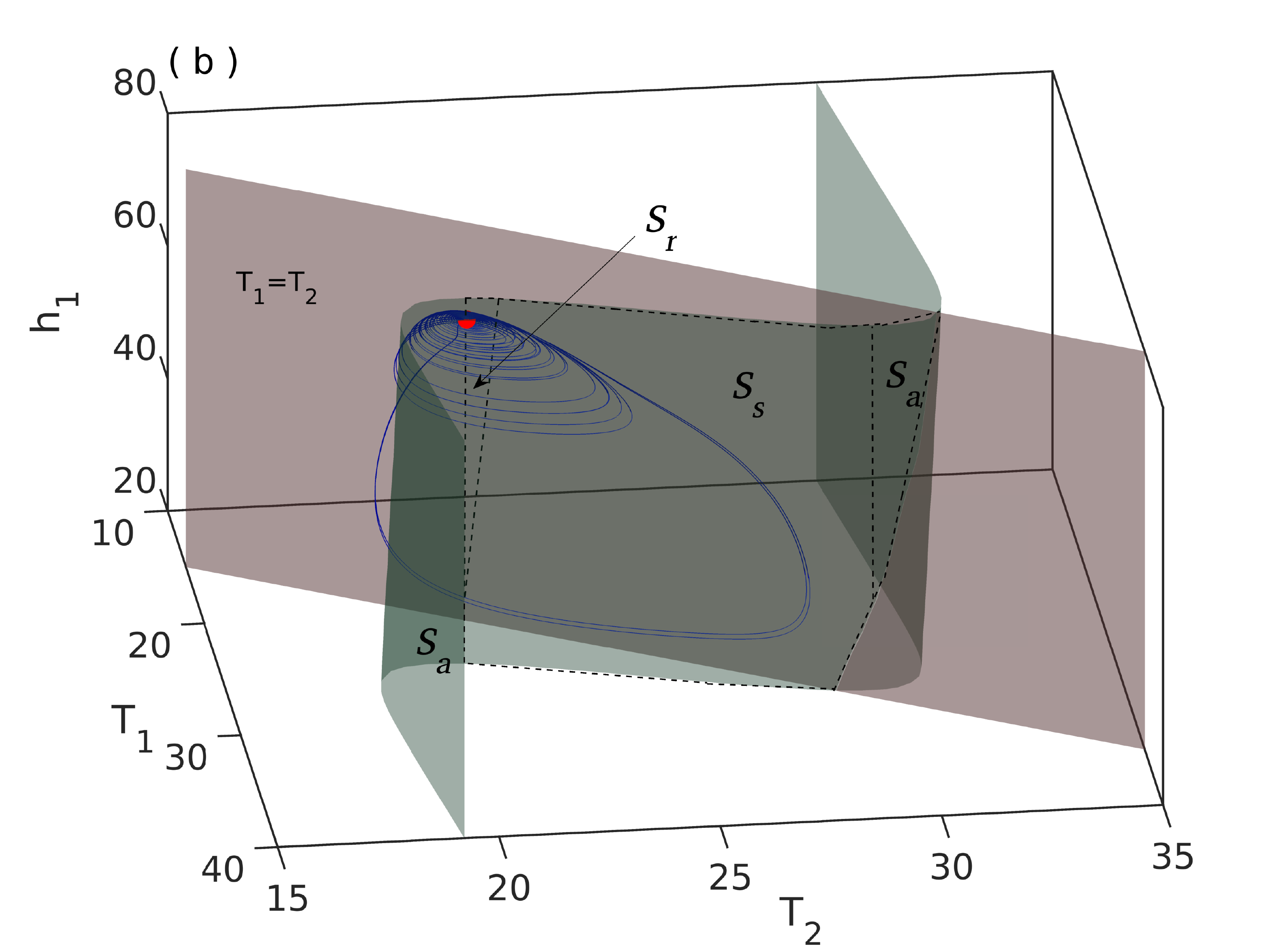}}
	\caption{Critical manifold $\mathcal{S}$ (green surface).  $\mathcal{S}_a$, attracting part; $\mathcal{S}_r$, repelling part; and $\mathcal{S}_s$, saddle part of the critical manifold. $T_1=T_2$ plane in brown, and a red circle indicates the position of saddle focus $(76.43, 27.28, 20.33)$. (a) Bounded chaotic attractor (blue line) for $\epsilon$=0.0984 (precrisis) and (b) expanded attractor (blue line) for extreme events $\epsilon$=0.0985 (postcrisis).}\label{figure_3}
\end{figure*}
\par The Jacobian $J_\mathcal{S}$ 
of the truncated system  is now derived at each point on the critical manifold $\mathcal{S}$ defined by using Eq.\ (\ref{eq.8}). The Jacobian
$J_\mathcal{S}$=$\begin{bmatrix} 
J_{ij} 
\end{bmatrix}_{2\times2}$, 
where 
\begin{widetext}
	\begin{equation}
	\begin{array}{lll}\label{eq.9}
	J_{11}=-\alpha+2\epsilon\mu(T_{2}-T_{1});\hspace{50pt} J_{12}=-2\epsilon\mu(T_{2}-T_{1});\\
	J_{21}=-\zeta\mu\Bigg(T_{2}-T_{r}+\dfrac{1}{2}(T_{r}-T_{r0})\bigg[1-\tanh\frac{\Big(H+h_{1}+\frac{bL\mu(T_{2}-T_{1})}{\beta}-z_{0}\Big)}{h_*}\bigg]\Bigg)+\\\hspace{30pt}\zeta\mu(T_{2}-T_{1})\Bigg(\dfrac{bL\mu}{2h_*\beta}(T_{r}-T_{r0})\bigg[\mbox{sech}^2\frac{\Big(H+h_{1}+\frac{bL\mu(T_{2}-T_{1})}{\beta}-z_{0}\Big)}{h_*}\bigg]\Bigg);\\
	J_{22}=-\alpha+\zeta\mu\Bigg(T_{2}-T_{r}+\dfrac{1}{2}(T_{r}-T_{r0})\bigg[1-\tanh\frac{\Big(H+h_{1}+\frac{bL\mu(T_{2}-T_{1})}{\beta}-z_{0}\Big)}{h_*}\bigg]\Bigg)+\\\hspace{30pt}\zeta\mu(T_{2}-T_{1})\Bigg(1-\dfrac{bL\mu}{2h_*\beta}(T_{r}-T_{r0})\bigg[\mbox{sech}^2\frac{\Big(H+h_{1}+\frac{bL\mu(T_{2}-T_{1})}{\beta}-z_{0}\Big)}{h_*}\bigg]\Bigg).
	\end{array}
	\end{equation}
\end{widetext}

\par We plot the critical manifold \cite{jg} in a 3D surface in Fig.\ \ref{figure_3} (green) and mark separate sections (dashed lines) of attracting ($\mathcal{S}_a$), repelling ($\mathcal{S}_r$) and saddle ($\mathcal{S}_s$) regions of the critical manifold using the sign of the eigenvalues of  $J_\mathcal{S}$ at each point of the surface. The saddle focus  $( 76.43, 27.28, 20.33)$ (red circle) is lying on  the border of the attracting and the repelling regions of the critical manifold. Starting from a {\it precrisis} point at $\epsilon=0.0984$, we obtain bounded chaos as  shown in  Figs.\ \ref{figure_1}(c), \ref{figure_1}(d), and \ref{figure_2}(d). Figure\ \ref{figure_3}(a) shows the trajectory of bounded chaos in 3D space traveling far away from the saddle focus, rather revolving around the triangular shaped repelling region ($\mathcal{S}_r$) of  the critical manifold. In contrast, Fig.\ \ref{figure_3}(b) shows a {\it postcrisis} chaotic attractor on the critical manifold $\mathcal{S}$ for $\epsilon=0.0985$. This is obtained by tuning the $\epsilon$ parameter near the critical point when the  trajectory is destined to travel a close vicinity of the saddle focus (red circle) in the repelling region. It takes longer time to spiral out since the trajectory gets slower near  the saddle focus and makes many small-amplitude oscillations before moving out for a global excursion, but it is reinjected to reach another close proximity of the saddle focus along the stable eigendirection. This is exactly the way a homoclinic chaos originates. However, in contrast to homoclinic chaos,  the trajectory is no more globally stable rather traces different trajectories during the re-injection period due to the presence of a global instability in the form of a channel-like structure. Thereby it originates large spikes with a variability of amplitude and  return time intervals that depend upon the number of small oscillations.

\begin{figure*}[ht]
	\centerline{\includegraphics[scale=0.25]{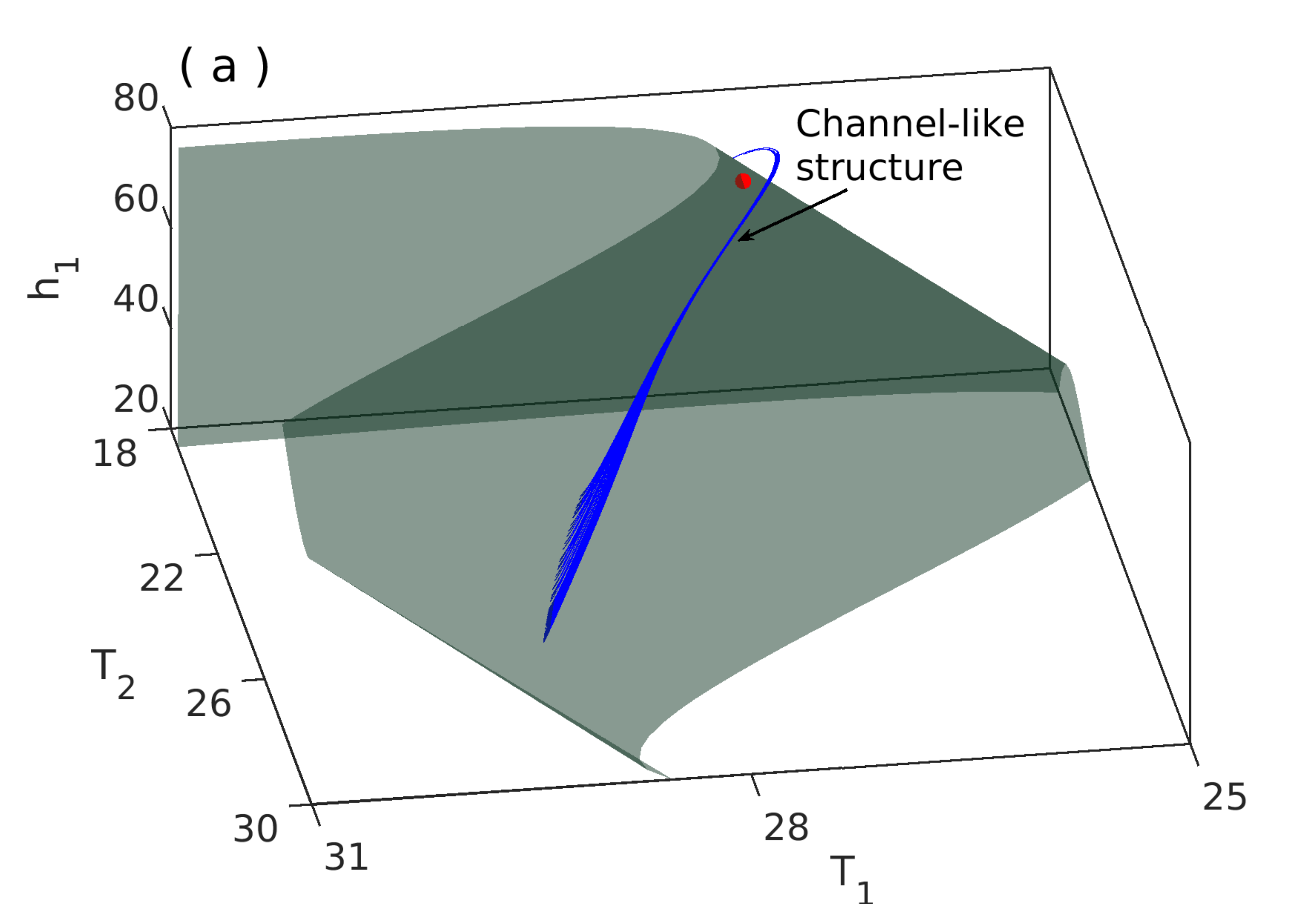}
		\includegraphics[scale=0.25]{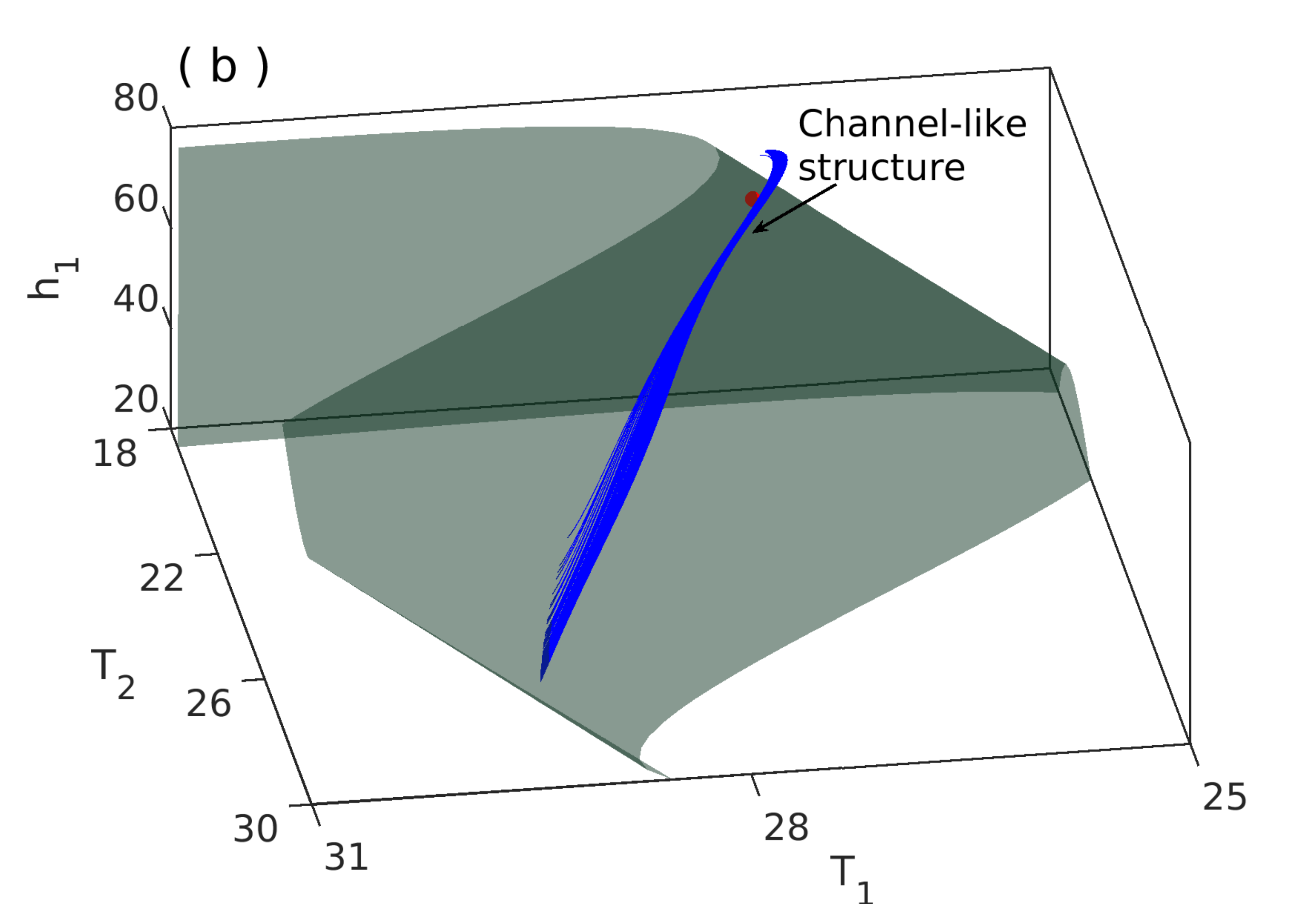}}
	\caption{Variation of the channel structure. Segments of a trajectory (blue lines) plotted for (a) $\epsilon$=0.0985 (a narrow channel) and (b) $\epsilon$=0.1076 (a wider channel) on the critical manifold $\mathcal{S}$ (green surface). Red circles are the positions of  saddle focus $(76.43, 27.28, 20.33)$ for (a) and $(73.87, 27.23, 20.52)$ for (b).}\label{figure_4}
\end{figure*}
\par Now we explain here the role of the channel-like structure that creates a global instability. The geometric singular perturbation theory \cite{gspt} explains the existence of a locally invariant slow-manifold $\mathcal{S}(r)$ close  to the critical manifold $\mathcal{S}$ within a domain $O(r)$, where attracting or repelling properties of the critical manifold remain unchanged. This invariant manifold consists of stable (attracting) and unstable (repelling) slow manifolds \cite{mmo}. As usual a slow increase in the size of a bounded chaotic attractor  occurs with the increase of $\epsilon$ and it enters the repelling part of the critical manifold at a critical value of $\epsilon$. After entering the repelling section, the trajectory tends to move toward the saddle focus along the stable manifold as mentioned above. When the trajectory reaches a close vicinity of the saddle focus, it spirals out along the 2D unstable manifold and attempts a large excursion to originate a large spike, but passes through a channel-like structure. The global stability of homoclinic chaos with a fixed global trajectory, as expected in such a situation, is broken by the channel-like structure. This structure appears due to changes in the alignment of the manifolds in  phase space as the control parameter $\epsilon$ is varied \cite{rajat2,rajat1,ee_prx}. To illustrate the influence of this channel-like structure, we follow the trajectory microscopically. We project collected data on a  segment of the trajectory as shown in Fig. \ref{figure_4}(a) from a long time series of large events shown in Fig.~\ref{figure_1}(e). The trajectory while spiraling out, passes through a very narrow channel in phase space. The system trajectory escapes from its small-amplitude oscillation while spiraling near the saddle focus and passes through this leaky channel. 

\par This instability region explicitly prescribes the location of the leak in  phase space and most importantly, its volume is inversely proportional to the rarity of large spiking events. The trajectory spends longer time in the narrow channel when the number of small oscillations are larger and, thereby it creates longer interval of the large events and also a wide spreading in the reinjection path of the trajectory that makes an amplitude variation of the large spikes or events. The role of the channel is further illustrated by increasing  $\epsilon=0.1076$ when the volume of the channel  increases in Fig.~\ref{figure_4}(b) plotted from a time series in Fig.~\ref{figure_1}(g) of chaotic MMO  (in the case of El Ni\~no situation). In this case, the trajectory spends less time in the channel and hence large spiking events are more frequent with low variability of amplitude. To summarize, a  collision  between the PD cascade from the left side and PA cascade from the right side of the system parameter ($\epsilon$) occurs at a crisis point \cite{fan} that makes a sudden change in the size of attractor. An additional  global instability of trajectories is  created by a channel-like structure in phase space.  The channel volume is modulated by the variation of $\epsilon$ and as a result, two kinds of events, the  extreme  events and the typical El Ni\~no events, evolve as shown in Figs. \ref{figure_1}(e) and  \ref{figure_1}(g), respectively, for two different values of $\epsilon$, one with large variability in amplitude and return time making its rare occurrence and another one with almost identical heights and more frequent occurrence of events.
\section{Extreme Events: Statistical properties}\label{stat}
We explore here the statistical properties of the extreme events, the probability distributions of event heights and interevent intervals. At first, a mean excess function is defined to identify a threshold level for qualifying extreme events.
\subsection{Mean excess function: Extreme event qualifier} 
We use a mean excess function \cite{cole,mean_excess} to define a threshold of an event. Any event larger than this threshold is declared as extreme. For a random variable $X$, it is defined as
\begin{equation}\label{eq.10}
\begin{array}{lcl}
e(u)=E(X-u~|~X>u),
\end{array}
\end{equation}
where $E(X)$ denotes expectation of the random variable $X$ and $u$ is the threshold value which varies between the minimum and maximum values of the observed data. This function gives an expected value of excess of a random variable over a certain threshold. Then we plot the function $e(u)$ with varying  $u$ and there we estimate a threshold value $u^*$ until  this functional relation  is a  better fit to a straight line. 
If any event crosses $u^*$, then the event is considered as an extreme event.  The extreme event qualifier threshold value is also obtained by considering few times more standard deviation than the mean value \cite{masolar2,njp2019}, deduced by using Peak Over Threshold approach \cite{massel}.
\begin{figure}[ht]
	\centerline{\includegraphics[scale=0.4]{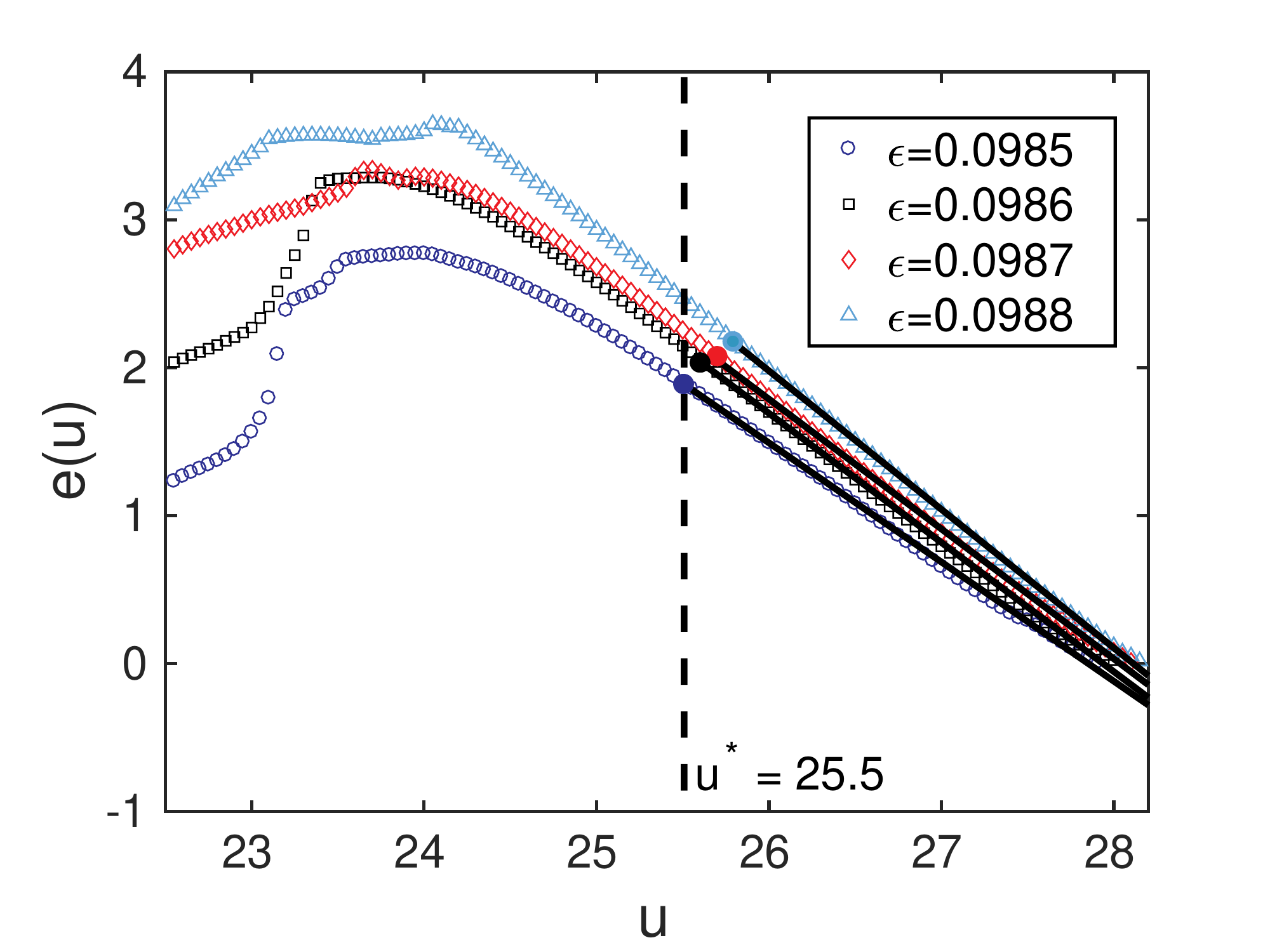}}
	\caption{Mean excess function plot against $u$ for different values of $\epsilon$:  $\epsilon=0.0985$ (blue circle), $\epsilon=0.0986$ (black square), $\epsilon=0.0987$ (red diamond), and $\epsilon=0.0988$ (cyan triangle). Filled circles represent the threshold values for different values of $\epsilon$.  Vertical dashed line is the threshold value $u^*=25.5$ for $\epsilon=0.0985$. A portion of curve is fitted by black solid line. }\label{figure_5}
\end{figure}
\begin{figure*}[ht]
	\centerline{\includegraphics[scale=0.4]{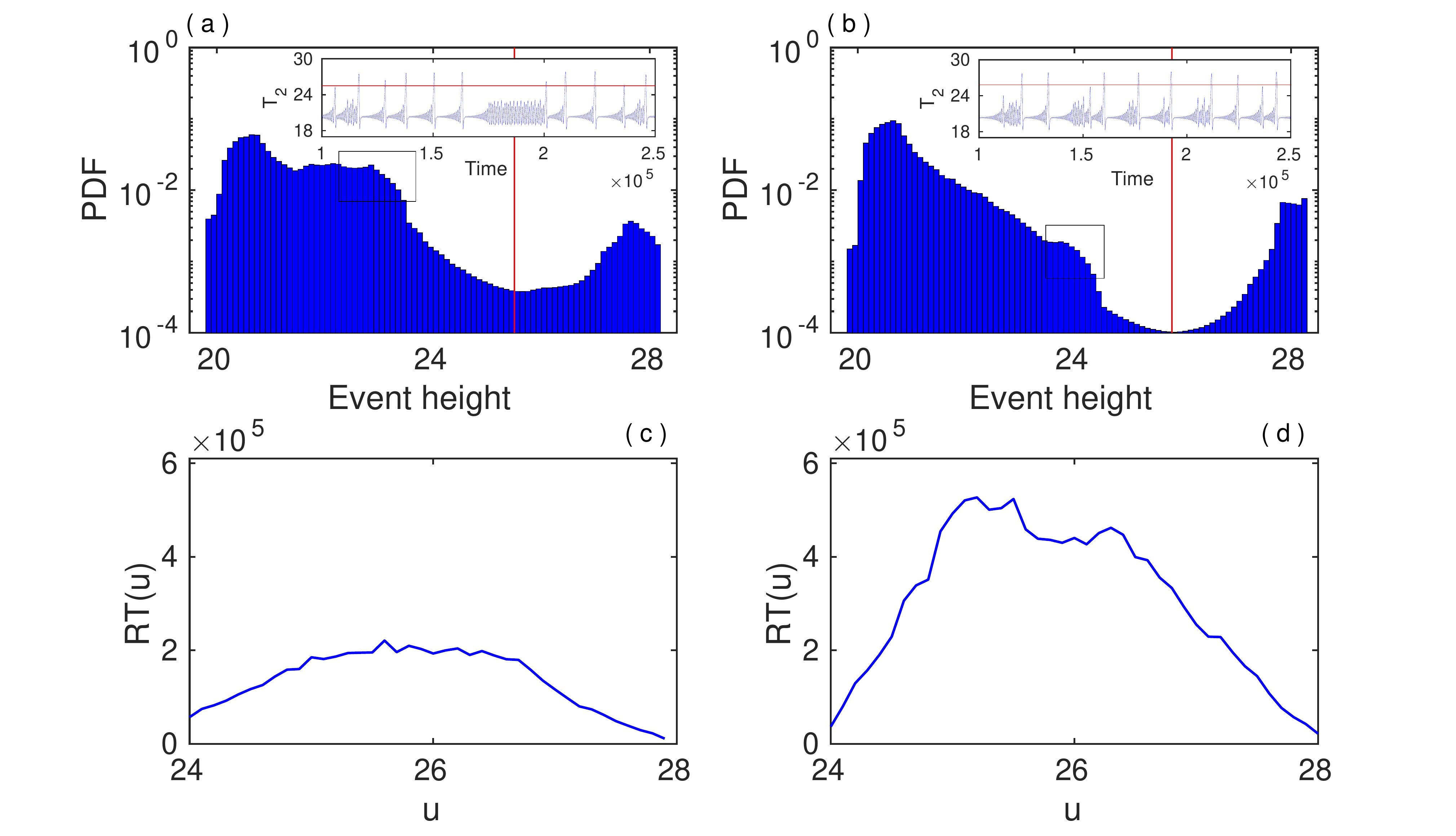}}
	\caption{Top: Probability density function of the event height in semi{-}log scale, inset figure shows the corresponding time series. Bottom: Corresponding return times: [(a) and (c)] $\epsilon=0.0985$ and [(b) and (d)] $\epsilon=0.0988$.}\label{figure_6}
\end{figure*}

\par 
As an example, for $\epsilon=0.0985$, in our system, the mean excess function $e(u)$ increases monotonically (blue line) with decreasing $u$ and reaches a peak, and finally decreases as seen in Fig.~\ref{figure_5}. However, a best straight line fit (black line) is possible until  $u=u^*=25.5$, beyond which the plot deviates slowly in the beginning and then largely (open blue circles) for lower $u$ values from the straight line. We mark $u^*=25.5$ as an extreme event qualifier (solid blue circle). Similarly, we obtain extreme event qualifiers  as $u^*=25.6$ (solid black circle), $25.7$ (solid red circle), and $25.8$ (solid cyan circle), respectively, for  $\epsilon=0.0986, 0.0987$, and $0.0988$. It is noted that the threshold value $u^*$  increases (shifts to the right) with increasing  $\epsilon$ and the rarity of occurrences of extreme events gradually decreases. The slopes of  straight line fits are $-0.8053$ (blue circle), $-0.8730$ (black square), $-0.8774$ (red diamond) and $-0.9362$ (cyan triangle) for $\epsilon=0.0985, 0.0986, 0.0987$, and $0.0988$, respectively. These slopes are used for approximating the values of shape parameter of the respective generalized Pareto distribution and to identify the nature of distribution as discussed  in the next section.

\subsection{Probability distribution of event heights}
\par  Now, probability density functions of all the peaks in a long run of time evolution of $T_2$ are drawn in semi-log scale in Figs.\ \ref{figure_6}(a) and \ref{figure_6}(b)  for two different values of $\epsilon$, {\it i.e.}, $\epsilon=0.0985$, very close to the interior crisis point and $\epsilon=0.0988$, slightly away from the crisis point. Simulations are run for sufficiently long time ($1.0\times10^{12}$ iterations) so that the distributions are saturated. Vertical lines (red) indicate the threshold level of extreme events estimated by the mean excess function (Fig.\ \ref{figure_5}). Clearly, both PDFs show multimodal non-Gaussian  distributions with extreme events larger than the threshold level.
Figure\ \ref{figure_6}(a) shows two dominant modes with an additional less dominant mode (marked by a rectangular box) for $\epsilon=0.0985$. The increasing values of $\epsilon$ push the additional mode to the right (rectangular box) and PDF approaches a bimodal distribution as shown in Fig.\ \ref{figure_6}(b) at $\epsilon=0.0988$. Insets show corresponding time evolution of $T_2$.  One mode corresponds to small amplitude oscillations and another mode corresponds to large amplitude oscillations.  Both the time signals of $T_2$ (insets) show an alternate large-amplitude spiking oscillations (related to extreme events) in alternate time sequence of spiraling small-amplitude oscillations. However, the height and return of large events show wide variability.
\par Once the intermittent large events are identified as extreme from a simulated time series, which are above a threshold $u^*$.  Now we introduce a function which is capable to capture the spatial status as well as temporal fluctuation of occurrence of the events. For this, we estimate the return time by taking an average of time-intervals of all the events at certain height, say, $u$,  and define it by $RT(u)$,
\begin{equation}
\begin{array}{l}\label{eq.rt}
RT(u)=\frac{1}{N}\sum\limits_{i=1}^{N}\text{IEI}_i(u),
\end{array}
\end{equation}
where $\text{IEI}_i(u)$ is the $i$-th interevent interval corresponding to all the events of the fixed height $u$, and $N$ is the number of interevent intervals collected from a long time series. Figure\ \ref{figure_6} (lower panel) shows the return time with respect to the event height $u$ for two exemplary values of $\epsilon$ used in the upper panels. We notice that the  PDFs in Figs.\ \ref{figure_6}(a) and \ref{figure_6}(b) and corresponding return times in Figs.\ \ref{figure_6}(c) and \ref{figure_6}(d) are inversely proportional in nature in the interval $24\le u \le 28$. So a larger return time of a particular event $u$ indicates that extreme events occur rarely. For both the cases, the return time of events increases with decreasing probability of number of events.
\begin{figure*}[ht]
	\centerline{\includegraphics[scale=0.60]{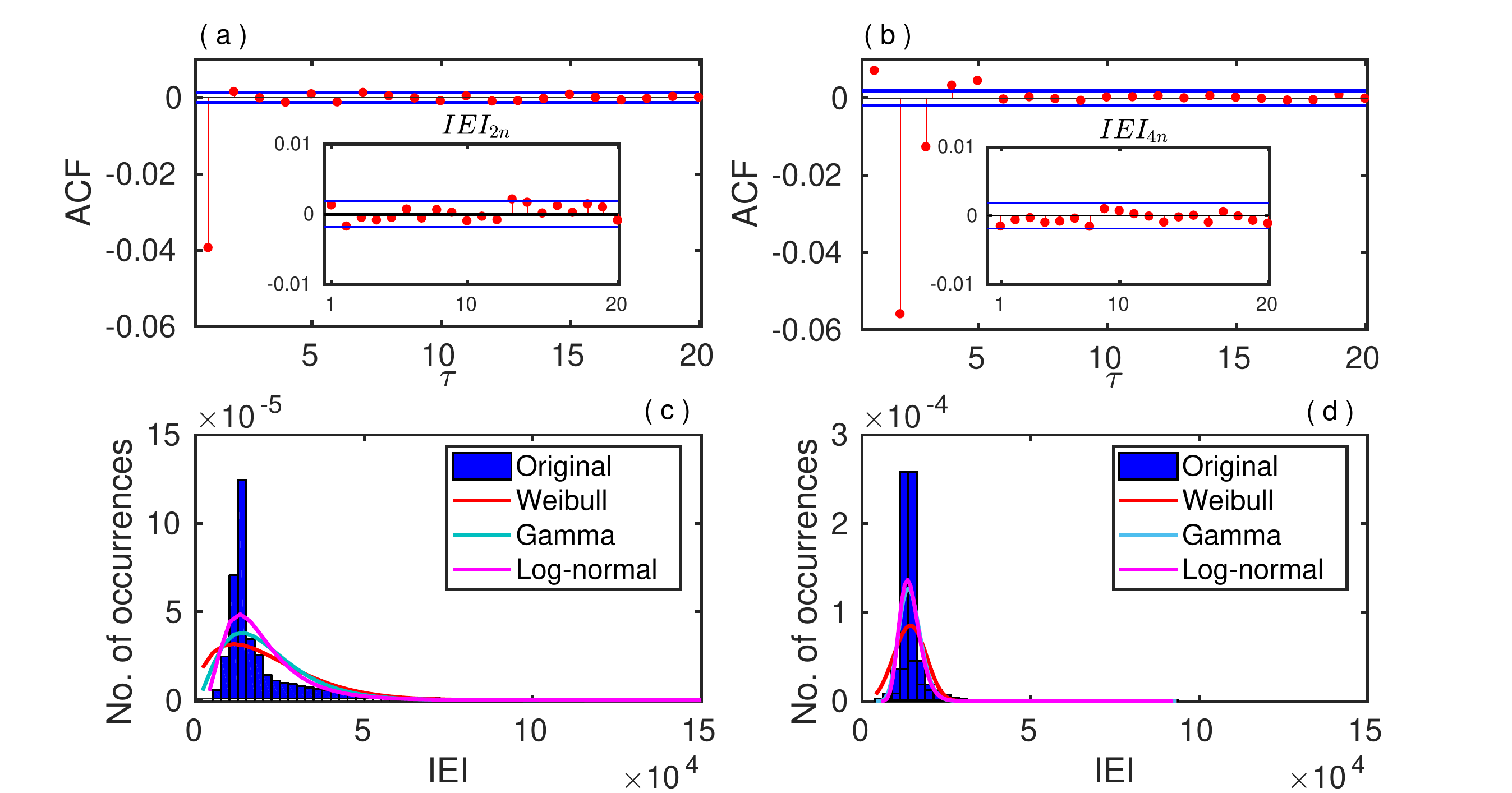}}
	\caption{Auto-correlation function (ACF) with varying lag parameter ($\tau$) for (a) $\epsilon=0.0985$, and (b) $\epsilon=0.0988$. Insets show  variation of ACF for the processes \{$IEI_{2n}$\} and \{$IEI_{4n}$\} for $n=1, 2, 3, \ldots$. PDFs of IEI are plotted (blue bars) in (c) $\epsilon=0.0985$, and (d) $\epsilon=0.0988$. Weibull (red line), Gamma (cyan line), and Log-normal distributions (magenta line) of events are fitted  to compare with original data (blue bars).}\label{figure_7}
\end{figure*}
\par Extreme events occur due to dynamical instability, which are responsible for the nonuniform behavior of the tail of the distribution. PDFs have different characteristics in different parameter regime and the tail of the distribution cannot be easily fitted with the well-known statistical functions \cite{sapsis}. We attempt here to find the characteristic feature of the tail of our observed PDFs in Figs.\ \ref{figure_6}(a) and \ref{figure_6}(b) with the help of the shape parameter in a generalized Pareto distribution (GPD) \cite{cole}. The probability density function of the Pareto distribution is 
\begin{equation}
\begin{array}{l}\label{eq.gpd}
G_{\xi,\gamma}(x)=
\begin{cases}
\frac{1}{\gamma}\big(1+\frac{\xi x}{\gamma}\big)^{-\frac{1}{\xi}-1}, & \text{$\xi\ne0$} \\
\frac{1}{\gamma}\exp(-\frac{x}{\gamma}), & \xi=0
\end{cases}
\end{array}
\end{equation}
where $x\ge0$ when $\xi$ is nonnegative, and $0\le x\le-\frac{\gamma}{\xi}$ otherwise.
Here $\gamma>0$ and $\xi$ respectively, are the scale and shape parameters of the distribution. The sign of the shape parameter $\xi$ delineates  the nature of the  distribution at the tail. In particular, if $\xi\ge0$, then the distribution has no upper limit, while $\xi<0$ implies the distribution has finite end-point. In our example, in Fig.\ \ref{figure_5}, we try to fit the mean excess function $e(u)$ against $u$ curve by a straight line, when we can express the mean excess function \cite{mean_excess} for the case of GPD as 
\begin{equation}
\begin{array}{l}\label{eq.eu}
e(u)\approx\frac{\gamma}{1-\xi}+\frac{\xi}{1-\xi}\rm{u},
\end{array}
\end{equation}
where $\frac{\xi}{1-\xi}$ is the slope of the best fitted straight line. For parameter values $\epsilon=0.0985$ and $0.0988$, the numerically calculated slopes of the straight lines in Fig.\ \ref{figure_5}, are $-0.8053$ and $-0.9362$, respectively. We obtain the shape parameters of the two distributions approximately as $-4.1361$ and $-14.674$, respectively, where the signs of the shape parameter $\xi$, tell us that the distributions have upper-bounded tails, which are shown in Figs.\ \ref{figure_6}(a) and \ref{figure_6}(b).

\subsection{Probability distribution of interevent interval}
\par  We study here the distribution and dependence structure of interevent interval (IEI).  Figures\ \ref{figure_7}(a) and \ref{figure_7}(b) depict that the autocorrelation functions (ACF) \cite{hamilton} of IEI   are out of small band ({\it i.e.}, significant, taking $95\%$ confidence interval) only for the first lag  at $\epsilon=0.0985$ and for the first three lags at $\epsilon=0.0988$, respectively (in this case, the values of ACF for fourth and fifth lags are very close to the confidence limit). These observations conclude that occurrence of extreme events are  correlated (in the short range). We have observed similar phenomena by moving the windows of smaller size through out the data. We thus identify \cite{brockwell} that the IEI process is stationary and can be modeled by ARIMA processes. 
From there, it can also be said that the process $\{IEI_{2n}:\ n=1, 2, 3, \ldots\}$ is an uncorrelated process [see the inset figure in Fig.\ \ref{figure_7}(a)] for the first case. Also the process $\{IEI_{4n}:\ n=1, 2, 3, \ldots\}$ is an uncorrelated process [see the inset figure in Fig.\ \ref{figure_7}(b)] for the second case. The similar comment is applicable for the process  $\{IEI_{2n-1}: \ n=1, 2, 3, \ldots\}$ [similarly to that of Fig.\ \ref{figure_7}(a)] as well as $\{IEI_{4n-1}, IEI_{4n-2}, IEI_{4n-3}:\ n=1, 2, 3, \ldots\}$ [similarly to that of Fig.\ \ref{figure_7}(b)]. We then use the Kolmogorov-Smirnov (KS) test \cite{pratt} to check whether the distributions of corresponding elements of these two sets  $\{IEI_{2n-1}, IEI_{2n}: \ n=1, 2, 3, \ldots\}$ and $\{IEI_{4n}, IEI_{4n-1}, IEI_{4n-2}, IEI_{4n-3}:\ n=1, 2, 3, \ldots\}$ are same or not. The $p$-values corresponding to $\epsilon=0.0985$ is $0.5792$. For $\epsilon=0.0988$, we get six $p$-values for six pairs $(IEI_{4n}, IEI_{4n-1})$, $(IEI_{4n}, IEI_{4n-2})$, $(IEI_{4n}, IEI_{4n-3})$, $(IEI_{4n-1}, IEI_{4n-2})$, $(IEI_{4n-1}, IEI_{4n-3})$, and $(IEI_{4n-2}, IEI_{4n-3})$  are $0.8464$, $0.6683$, $0.5478$, $0.5039$, $0.479$, and $0.7091$, respectively. Here, $p$-value denotes the probability of obtaining test results at least as extreme as that of the observed one(s)
under the null hypothesis. Since the $p$-values are very high, we fail to reject the null hypothesis of equality of any of these pairs of distributions. The KS test also confirms that the distributions of all the processes are also same as that of the whole data of IEI. For different values of $\epsilon$, the processes can be thought as a renewal process (of events) \cite{mitov} or more generally as a point process.
\par We now try three known parametric family of distributions (Weibull, Gamma, and Log-normal distributions) to fit the histogram of IEI (data collected from our numerical experiment) for two different aforementioned values of $\epsilon$ in Figs.\ \ref{figure_7}(c) and \ref{figure_7}(d) to help us understand the nature of occurrence of next extreme events. We have also observed that Log-normal distribution is the best fit for both the cases. In Appendix B, we write the expressions of PDFs of IEI and present a chart of the estimated parameters of these distributions (Table I).   
\section{Predictability of extreme events} \label{predict}
The dependence structures of IEI are observed in Figs.\ \ref{figure_7}(a) and \ref{figure_7}(b) which indicate about predictability of extreme events \cite{holgar}.
For predictability, we use ARIMA model \cite{chatfield}, histogram of the IEIs and a corresponding box plot analysis \cite{box}. A preliminary description related to the predictive property of box plot analysis is described in Appendix C.

\par
\begin{figure}[ht]
	\centerline{\includegraphics[scale=0.27]{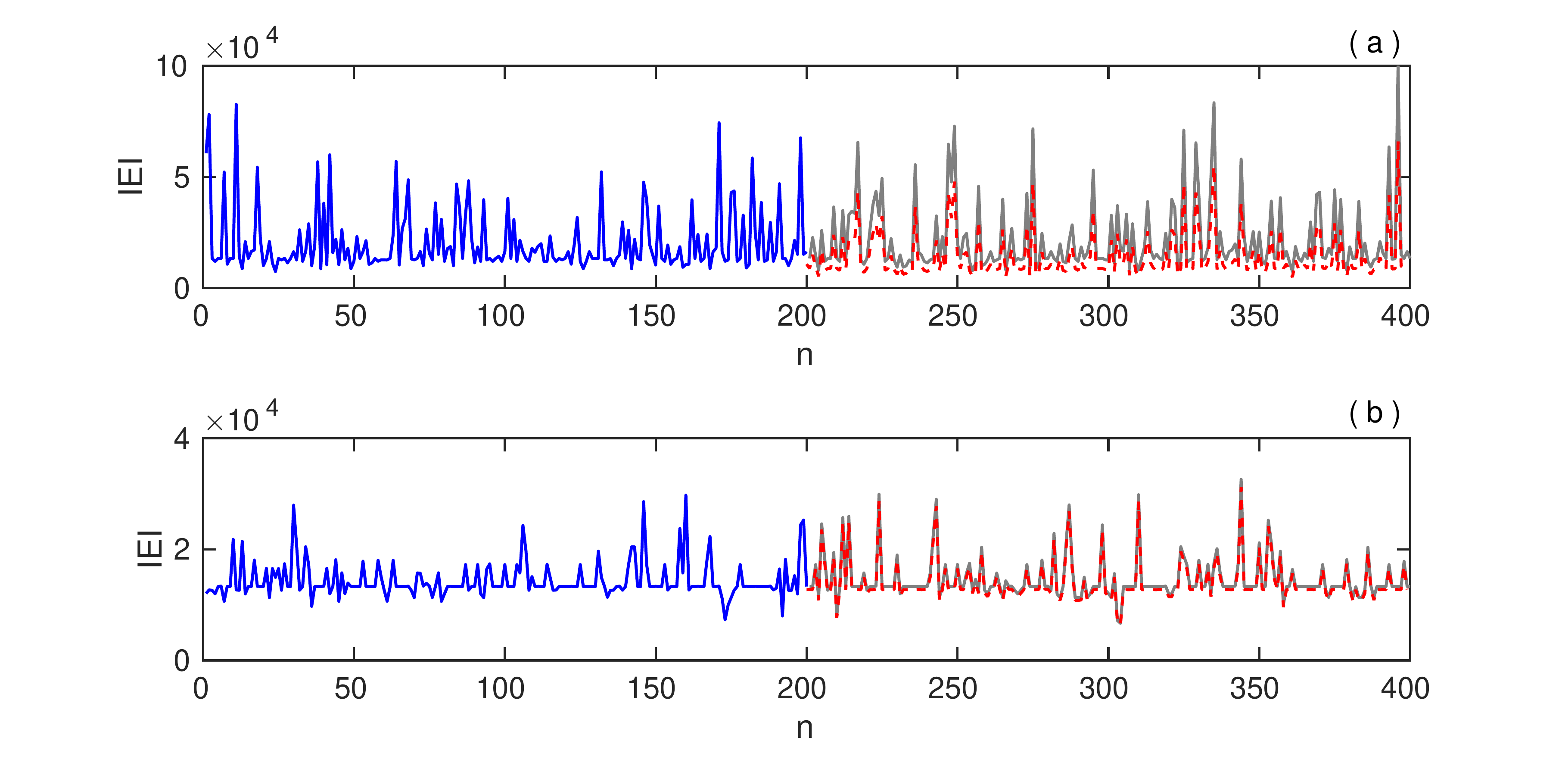}}
	\caption{Prediction plots of one-step-ahead forecast using ARIMA $(1,0,0)$ time-series model based on the simulated IEI's for (a) $\epsilon=0.0985$ and (b) $\epsilon=0.0988$. Blue solid curves: simulated IEIs used for forecasting; red dash curves: predicted IEIs; and gray solid curve: exact simulated IEIs. }\label{fig_8}
\end{figure}

  ARIMA time-series forecasting model is used to fit the data at first. For this, we find that the best fitted model for data is ARIMA $(1, 0, 0)$, {\it i.e.}, a model with only autoregressive part of lag one and no integrated or moving average part. We use  half of the 400 samples to find a best fit of the model and on the remaining half we use the one-step-ahead prediction using the above AR(1), or, equivalently, the ARIMA $(1, 0, 0)$ model.  We plot the iteration of our simulated IEIs (blue solid curve) and predicted IEIs (red dash curve) for $\epsilon = 0.0985$ and $\epsilon = 0.0988$, respectively, in Figs.\ \ref{fig_8}(a) and \ref{fig_8}(b). For this predictability, we use one-step-ahead ARIMA $(1, 0, 0)$ time-series estimation. Figure\ \ref{fig_8} shows a deviation of the time of occurrences (plotted along the ordinate) whose abscissa is the index of the occurrences of IEI. From the mean absolute deviation (MAD) and the root mean standard error (RMSE), predictions are of the order of $10^4$ for Fig.\ \ref{fig_8}(a) and $10^3$ for Fig.~\ref{fig_8}(b). This is reasonable for predicting such extreme events, whose distribution of time intervals of occurrence, IEI, have a very wide range. Here we observe that if the value of $\epsilon$ is shifted away from critical crisis point, the predictability of extreme events becomes finer  because the chaotic fluctuation of time signal of $T_2$ decreases varying with parameter and it reaches to MMO, where fluctuation of temporal evolution of system drops significantly. As a result, it may quite easy to predict such  El Ni\~{n}o events using ARIMA method. We investigate further the question of predictability in a different way.
\begin{figure}[ht]
\centerline{\includegraphics[scale=0.36]{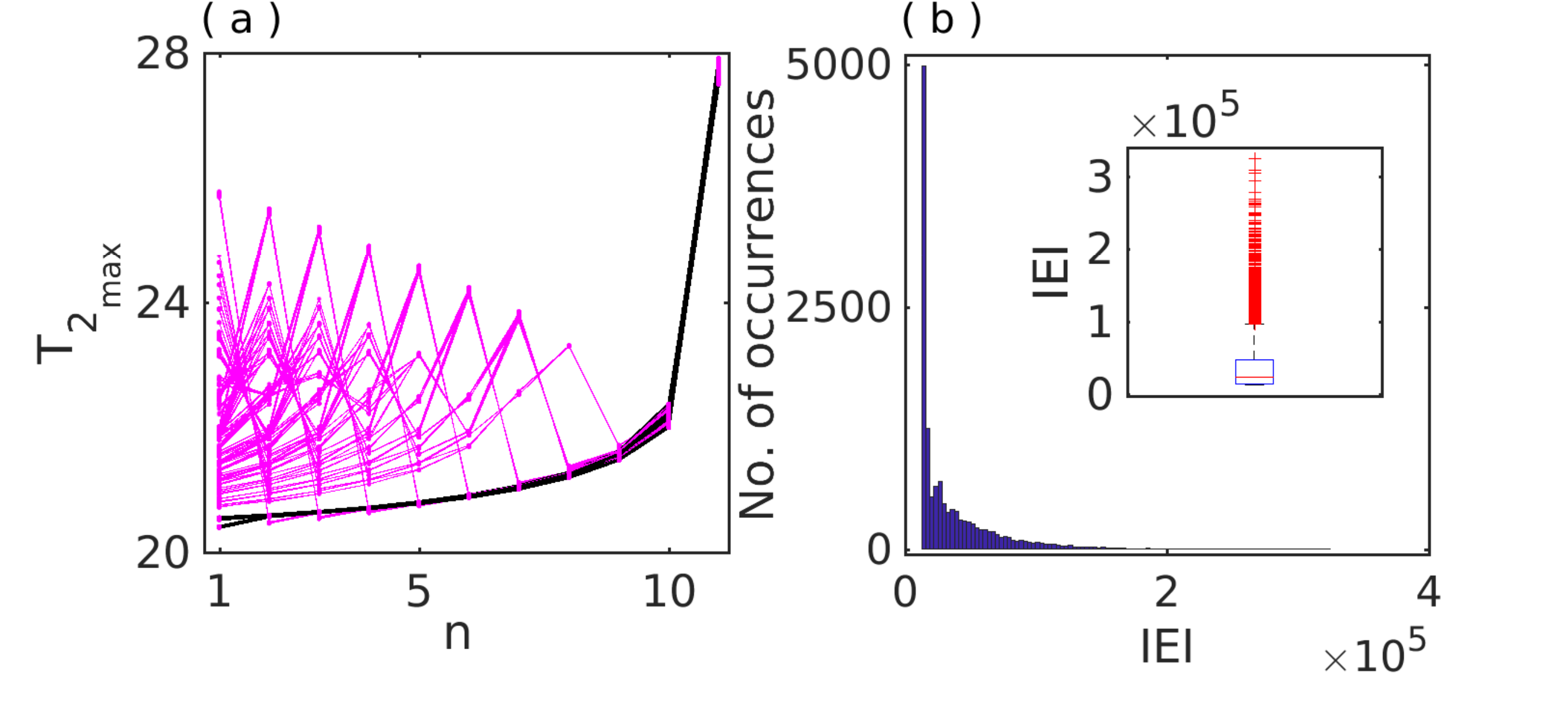}}
\caption{(a) Local maxima of $T_{2}$ before appearing the event height $27.5$ and above (10 last maxima) are plotted. Black and magenta curves are the events which occur through spiral outward and bounded chaotic oscillations, respectively. (b) Histogram of the interevent intervals at $\epsilon=0.0985$ {for all the events of height $27.5$ and above followed by spiral outward as well as bounded chaotic oscillations} and corresponding box plot in the inset.}\label{figure_7_b}
\end{figure}
\begin{figure*}[ht]
	\centerline{\includegraphics[scale=0.53]{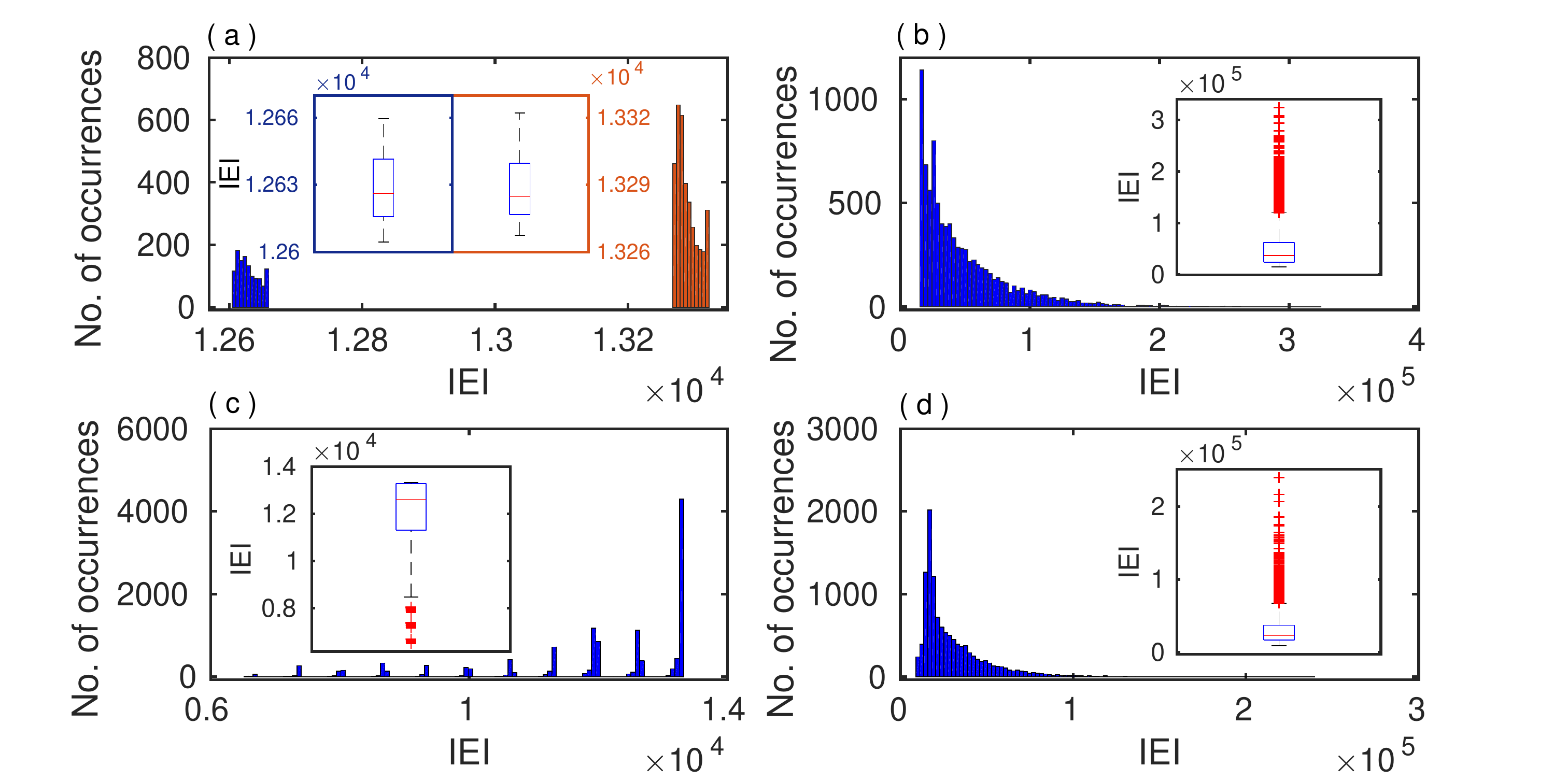}}
	\caption{Histogram of interevent intervals at $\epsilon=0.0985$ which are followed by spiral outward oscillations (left panel), and bounded chaos (right panel),  respectively. The upper [(a) and (b)] and lower [(c) and (d)] panels are respectively drawn by considering the events of height larger than $27.5$ and $25.5$. The inset figures show the corresponding box plots.}\label{figure_7_c}
\end{figure*}
\par A careful observation of the simulated time evolution of $T_2$ in Fig.~\ref{figure_1}(e) and the inset in Fig.~\ref{figure_6}(a) reveals that two types of spiking patterns appear, which may be characterized as extreme events. A class of large spiking events called here as extreme events emerge after spirally outward with small amplitude oscillations; another class of extreme events emerge after a relatively longer duration of bounded chaotic oscillations around the saddle focus. To distinguish these two types of extreme events, we plot a set of last $10$ local maxima of $T_{2}$ before occurring an event of height larger than or equal to $27.5$ as shown in Fig.\ \ref{figure_7_b}(a). Extreme events that emerge via two different processes, small amplitude oscillations  and bounded chaotic oscillation, are clearly distinguishable. The peak value of the spiraling oscillation ($T_{2_{max}}$) monotonically increases (black line) until they reach an event height larger than $T_{2_{max}}\ge27.5$. However, during the  emergence of large events from the bounded chaotic oscillation, there is no such increasing trend before the extreme events, rather a serpentine pattern (magenta line) is seen. Inspired from this fact, we separately investigate the question of predictability of extreme events arising out of the two cases.  
\par Figure \ref{figure_7_b}(b) presents a  histogram of IEI for all the events of height $27.5$ and above that includes for both the cases and it shows a unimodal long-tail distribution. A corresponding box plot (see Appendix C for details) is drawn in the inset, which we elaborate here for our results. Five measures for this box plot are estimated from our simulated time-series data:  $\mbox{{\it lower adjacent}}=12604.38$, ${\it first} {\it quartile} ~\mbox{Q}_1=13291.14$, $~\mbox{{\it median}}=22822.30$, ${\it third} {\it quartile} ~\mbox{Q}_3=46548.3$, and $\mbox{{\it upper adjacent}}=96434.04$. These measures tell us that once an event of height  $27.5$ and above occurs, then next such type of event will occur within the time intervals $[12604.38,13291.14]$, $[12604.38,22822.30]$, and $[12604.38,46548.3]$ with respective probabilities $0.25$, $0.5$, and $0.75$. It is almost sure that these particular type of events will occur within the interval $[12604.38,96434.04]$. Red-marks (+) in the box plot indicate seemingly the hard to predict extreme events which are about 6.55\% of them. From our results,  we conclude that the interquartile range (IQR)=$Q_3-Q_1=33257.16$, and approximately 6.55\% of the events are not predictable with this standard methodology, which in statistical terms may often be called {\it outliers}.

\par Next we investigate all extreme events of height $27.5$ and above on the basis of their emergence, either through spiral outward or bounded chaotic oscillations as classified above. Results are shown in Figs.\ \ref{figure_7_c}(a) and \ref{figure_7_c}(b) and measurements are given in the Table II (Appendix C). Figure\ \ref{figure_7_c}(a) depicts a multimodal distribution. We have drawn two different box plots for these two disjoint components of the distribution in the insets. Figures\ \ref{figure_7_c}(a) and \ref{figure_7_c}(b) together show the data described in Fig.\ \ref{figure_7_b}(b) may be suitably modeled by a mixed distribution with two major components, spiral outward and bounded chaos oscillations, with corresponding frequencies 35\% and 65\%, respectively. Within the spiral outward component, two smaller components are visible, one with frequency 9\% and another with 26\% with IQR$=25.862$ (left box) and IQR$=22.99$ (right box) which are significantly small compared to bounded chaos. For both these cases, we do not get any outliers which indicate better predictability of such events and the events occur through spiral outward oscillation is almost predictable. Figure\ \ref{figure_7_c}(b) shows an long-tail unimodal distribution for the interevent intervals of the extreme events of height $27.5$ and above occurring through bounded chaos.
\begin{figure*}[ht]
	\centerline{\includegraphics[scale=0.509]{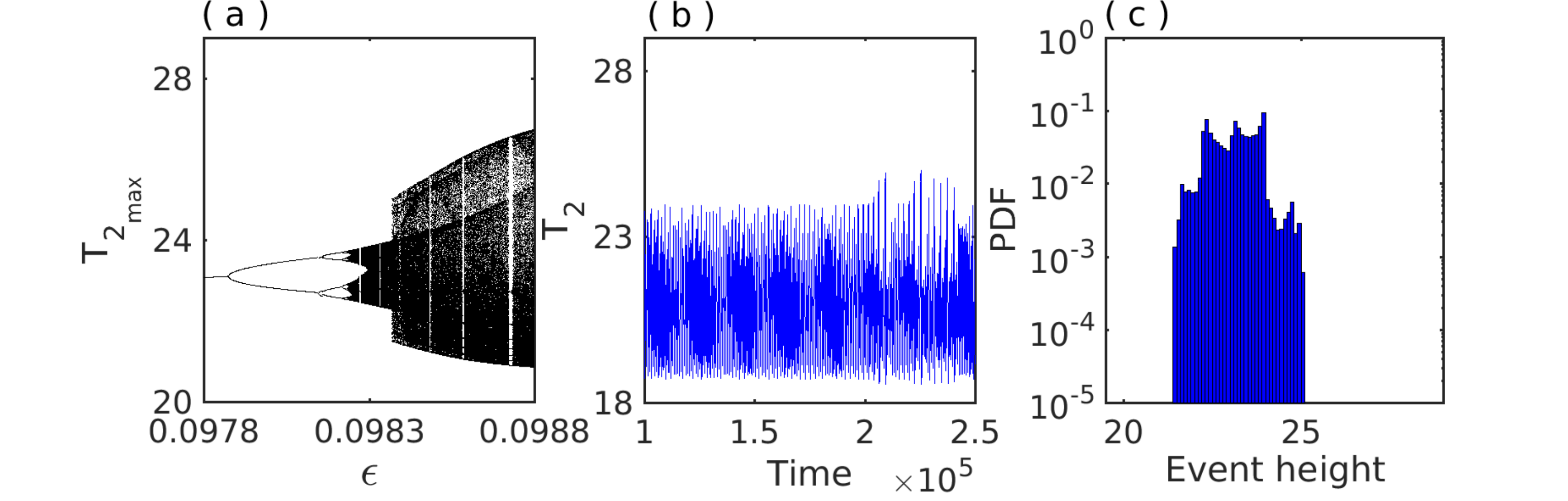}}
	\centerline{\includegraphics[scale=0.5]{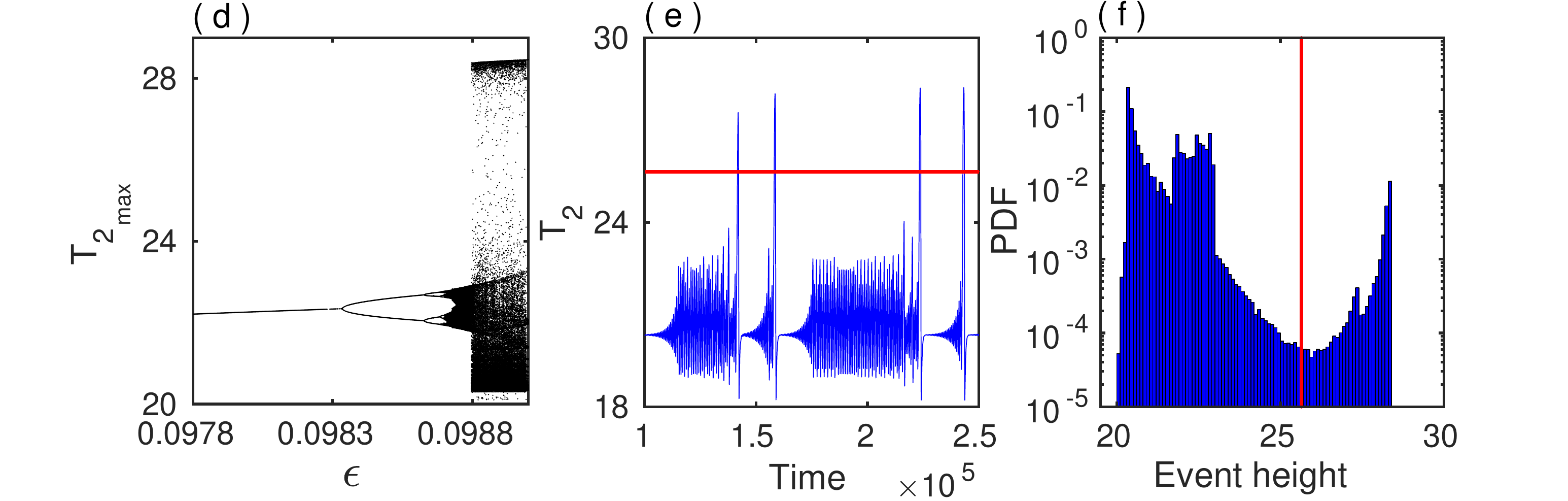}}
	\caption{Bifurcation diagram of the variable $T_{2}$ with respect to $\epsilon$ for (a) $r=\frac{1}{300}$ and (d) $r=\frac{1}{450}$. Temporal evolution of $T_2$ at (b)  $\epsilon=0.09837$, $r=\frac{1}{300}$, and (e) $\epsilon=0.0988$, $r=\frac{1}{450}$ and corresponding probability density functions are shown in (c) and (f), respectively. Red lines in (e) and (f) are the threshold values at $u^*=25.65$ (measured using mean excess function).}\label{figure_8}
\end{figure*}

\par A similar study is performed in the lower panel of Fig.\ \ref{figure_7_c} for all event heights (for which $T_{2_{max}}$ exceeds the predefined threshold $25.5$) instead of a specific height. For such cases, similarly we have separated out all the events which are occurring through spiral outward and bounded chaotic oscillations. Figures\ \ref{figure_7_c}(c) and \ref{figure_7_c}(d) respectively represent the distributions of IEI for these two cases. For the spiral outward induce extreme event which exceeds the length $25.5$, almost every interevent intervals are concentrated within a very small abscissa length. It shows the feature of periodicity for the occurrence of extreme event followed by small amplitude oscillations.
However, bounded chaos induces events that exhibit a unimodal long-tail distribution. 
The IQR and outlier for such events are respectively $1971.79$ and $5.74\%$, while $20139.18$ and $5.64\%$ for bounded chaos case. This study clearly indicates that the spiral outward induces events are more predictable than bounded chaos induces events.

\section{Effect of slow-fast timescale}\label{timescale}
\par The parameter $r$ controls the timescale of the slow variable of the system (\ref{eq.2}). It plays a crucial role in the onset of a sudden large expansion of the attractor from a bounded state and in the origin of extreme events. For  elaboration, we fix all the parameters as considered  above and draw bifurcation diagrams of $T_{2}$ for two additional choices of $r$. Figures\ \ref{figure_8}(a) and \ref{figure_8}(d) show two bifurcation diagrams against  $\epsilon$ for   $r=\frac{1}{300}$  and $r=\frac{1}{450}$, one larger and one smaller than  $r=\frac{1}{400}$.  We have already drawn the scenario of PD and PA cascade collision in Fig.\ \ref{figure_2}(d) in Sec.\ \ref{origin} for $r=\frac{1}{400}$.  The extreme event occurs at a postcrisis region after transition point $\epsilon \approx 0.09848$ in Fig.\ \ref{figure_2}(d).  For a faster timescale, $r=\frac{1}{300}$, an expansion of $T_{2_{max}}$  occurs at a transition point  $\epsilon\approx0.09837$, but it is not large enough to originate  extreme events, which is also confirmed by its temporal evolution in Fig.\ \ref{figure_8}(b) and an almost Gaussian probability distribution of event heights in Fig.\ \ref{figure_8}(c). On the other hand, if we consider a lower  $r=\frac{1}{450}$ value, then the critical  $\epsilon$ for the onset of a sudden large expansion shifts to a larger value, which is clearly observed in Fig.\ \ref{figure_8}(d) [cf. Fig.~\ref{figure_2}(d)]. The corresponding time series at $\epsilon=0.0988$ is plotted in Fig.~\ref{figure_8}(e) where the large events are really extreme events since they cross a threshold $u^*=25.65$ (horizontal red line), which is calculated by the mean excess function plot. PDF also changes significantly at $\epsilon=0.0988$ for a lower $r=\frac{1}{450}$ as show in Fig.\ \ref{figure_8}(f) clearly shows a multimodal non-Gaussian  distribution. So it is clear that timescale $r$ plays an important role for generating extreme events by shifting $\epsilon$ for the onset of extreme events. For a smaller value of $r$, {\it i.e.}, for a slower rate of change in $h_1$, a larger value of the strength of zonal advection $\epsilon$ is necessary to reach the crisis point where the enlargement of the attractor in phase space occurs leading to extreme events.

\section{Conclusion}\label{conclu}
\par A low-dimensional slow-fast  ENSO model has been investigated  to understand the dynamical origin of extreme and related El Ni\~no events. Using bifurcation diagrams of SST against a system parameter $\epsilon$ connected to the strength of difference in SST in the southern and eastern Pacific, we identified two different kinds of interesting events,  the typical chaotic MMO type El Ni\~no events  and another extreme events. El Ni\~no events emerge more frequently and almost in identical heights while the second type of events shows a wide variability in amplitude and interevent intervals. For a demonstration of both the events, besides drawing bifurcation diagrams of the system by varying the system parameter, we analytically derived a critical manifold of the system and located the regions of instabilities in phase space  of the system. We found a  critical value of the control parameter $\epsilon$, where a sudden expansion of the chaotic attractor was observed  due to an interior crisis. This interior crisis emerged due to a merging of PD and PA cascade of bifurcations at a critical point when the extreme events emerge by a self-induced switching between small-amplitude oscillations and large-amplitude events. In fact, a homoclinic chaos was expected, in such a situation, with a local instability of the saddle focus of the slow-fast system along with a global stability of the trajectory when large identical spiking events were usually seen alternately with randomly varying number of small oscillations. However, an additional  instability exists in phase space of the system in the form of a channel-like structure that prevents the global stability of the trajectory leading to large variation in amplitude and return time of the large spiking events. This variation in amplitude depends on the volume of this channel. For a  relatively narrow channel, a trajectory while spirally moving out of a saddle focus of the system, spent a longer period of time inside the channel and finally made a large global excursion to form an extreme event. The reinjected trajectories or the repeat global excursions were no more stable and irregular leading to a large variation of amplitude and return time of events. For a wider channel, a trajectory easily passed through it without spending much time inside it and hence the reinjected trajectories in the repeat global excursion were more stable to make almost identical height and more frequent chaotic MMO-like typical El Ni\~no events as reported earlier. We classify the large spiking events as extreme events, which are larger than an estimated a threshold defined by a mean-excess function. 
\par PDF of event heights showed a multimodal distribution which tends to be bimodal  with an increase in the system parameter $\epsilon$. For characterization of the occurrence of large events, PDF of interevent intervals has been fitted by Weibull, Gamma, and Log-normal distributions and out of the three, the Log-normal distribution is best fitted.  A dependence nature of the interevent intervals is obtained from ACF plot that illustrates possible implication of the predictability of (resulting) extreme events. We have discussed the predictability of extreme events, first using ARIMA models, and then in more details with box plot analysis. We show that the ARIMA models can predict the events with a standard error of the order of $10^4$. We then indulge in in-depth analysis via box plot. For our predictability purpose, we categorize each events in two different classes based on their emergence. One class of events takes much shorter time to occur, which originate via spiraling out from the saddle set and from a knowledge of such events and their occurrence, their return time is  predicted with a good accuracy; the second class of events on an average takes much longer time and they originate via bounded chaos where they spend more time before a large event. Our box plot analysis reveals, knowing that such events occur, predictability is found less accurate than the former one, because of the duration of the time spent in bounded chaotic motion. The slow-fast ratio parameter of the system plays a significant role in the  onset of  extreme events. While real data are available on chaotic MMO-like decadal emergence of extreme events, we are yet to find a real data set to verify the existence of our classified extreme events. However, we explain how and when two events, El Ni\~no events and extreme events may emerge in the low-dimensional slow-fast climate model against a system parameter variation, and explained the phenomena using a common dynamical mechanism, which has so far been missing. Furthermore, we tried to address the question of the predictability of extreme events.

\medskip

\par {\bf Acknowledgments:}
DG was supported by SERB-DST (Department of Science and Technology), Government of India (Project No.  EMR/2016/001039). SKD  acknowledges support by the Division of Mechanics, Lodz University of Technology, Poland. AR thanks Arindam Mishra for useful  discussions.\\

\section*{APPENDIX A} \label{appendix_a}
 We make a linear stability analysis of the equilibrium points of the system (\ref{eq.7}), which
 has one axial equilibrium point $(0,T_{r},T_{r})$ and one  interior equilibrium point $(h_{1}^*,T_{1}^*,T_{2}^*)$. The stability of the axial equilibrium  $(0,T_r,T_r)$ is obtained from the Jacobian matrix $J$ of the 3D system (\ref{eq.7}),
 \[
 J=
 \left[ {\begin{array}{ccc}
 	-r & \frac{bLr\mu}{2\beta} & -\frac{bLr\mu}{2\beta} \\
 	0 & -\alpha & 0 \\
 	0 & -p & -\alpha+p\\
 	\end{array} } \right],
 \]	
 where $p=\zeta\mu\frac{\rm{T_{r}}-T_{r_{0}}}{2}\big\{1-\tanh\frac{H-z_{0}}{h_*}\big\}$. Its eigenvalues are $-r,-\alpha, p-\alpha$. For our  choice of system parameters, the value of $p$ becomes $0.0141$, this yields $p>\alpha$ and hence the axial equilibrium  is a saddle point. Now we obtain,
 \[
 [J+rI]^m=
 \left[ {\begin{array}{ccc}
 	0 & (p-\alpha)^{m-1}\frac{bLr\mu}{2\beta} & -(p-\alpha)^{m-1}\frac{bLr\mu}{2\beta} \\
 	0 & -\alpha^m & 0 \\
 	0 & -\alpha^m-(p-\alpha)^m & (p-\alpha)^m\\
 	\end{array} } \right],
 \]
 when the generalized eigenvector of order $m$ corresponding to the  eigenvalue $-r$ is $[1~0~0]^{tr}$ (here $tr$ denotes the transpose of the matrix), for all $m\in\mathbb{N}$. For the other eigenvalue $-\alpha$, one can write
 \[
 [J+\alpha I]^m=
 \left[ {\begin{array}{ccc}
 	(\alpha-r)^m & \frac{(\alpha-r)^m-p^m}{\alpha-r-p}\frac{bLr\mu}{2\beta} & \frac{(\alpha-r)^m-p^m}{p+r-\alpha}\frac{bLr\mu}{2\beta} \\
 	0    & 						0									& 						0 \\
 	0    & 					-p^m	 								&					 	p^m\\
 	\end{array} } \right],
 \]
 when its eigenvector is $[0~1~1]^{tr}$.
 
\par The stable subspace of the saddle point $(0,T_r,T_r)$ is the eigenspace generated by the generalized eigenvector $\{[1~0~0]^{tr},[0~1~1]^{tr}\}$, which, in reality, is the $T_1=T_2$ plane.
\par The  $T_1=T_2$ plane is the invariant manifold of the system since when we have $\frac{d}{d t}[T_1-T_2]=0$ and, for all the choices of initial conditions, the system remains confined to this plane. The tangent space of the $T_1=T_2$ plane is $T_1=T_2$ itself, which is a stable subspace of system (7). In this stable manifold, our original 3D system can be reduced to a 2D system,
 \begin{equation}
 \begin{array}{lll}\label{eq.11}
 \dot{h}_{1}=-rh_{1},\\
 \dot{T}=-\alpha(T-T_{r}),
 \end{array}
 \end{equation}
 where $T_1=T_2=T$.   It is easy to check that the axial equilibrium point $(0, T_{r}, T_{r})$ is a stable node in this plane. Since our system is 3D and the stable manifold is 2D, so the unstable manifold is 1D, which is spanned by the generalized eigenvector $[\frac{2(\alpha-r-p)\beta}{bLr\mu}~0~1]$ corresponding to the positive eigenvalue $p-\alpha$ of $J$.
\par For ${T_1}\neq {T_2}$, the interior equilibrium point $(h_{1}^*, T_{1}^*, T_{2}^*)$ satisfies
\begin{widetext}
 	\begin{equation}
 	\begin{array}{l}\label{eq.12}
 	h_{1}^*=-\dfrac{bL\mu(T_{2}^*-T_{1}^*)}{2\beta},\;\;\;\;\;\;
 	 	\alpha(T_{1}^*-T_{r})+\epsilon\mu(T_{2}^{*}-T_{1}^{*})^2=0,\\
\dfrac{2[(\frac{\alpha}{\mu}+\zeta T_r)-(\epsilon+\zeta)T_2^*+\epsilon T_1^*]}{\zeta(T_r-T_{r0})}=1-\tanh\dfrac{\Big(H+\frac{bL\mu(T_{2}^*-T_{1}^*)}{2\beta}-z_{0}\Big)}{h_*}.
 	\end{array}
 	\end{equation}	
\end{widetext}

 \begin{figure}[ht]
	\centerline{\includegraphics[scale=0.42]{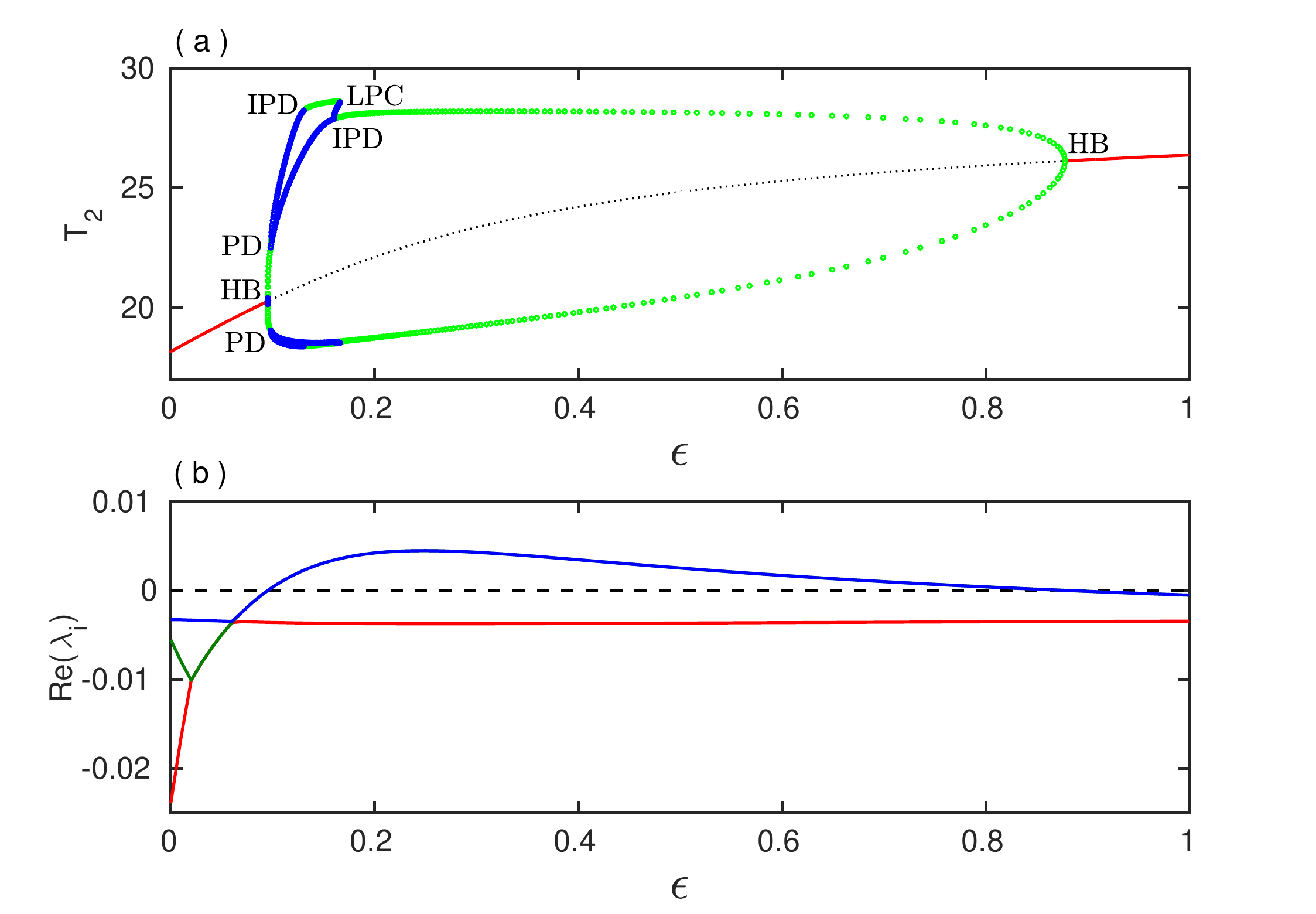}}
	\caption {(a) Bifurcation diagram of the interior equilibrium point $(h_{1}^*, T_{1}^*, T_{2}^*)$ with respect to $\epsilon$ by plotting the maxima of $T_2$. (b) Variation of the real parts of the eigenvalues for the interior equilibrium point by changing $\epsilon$. PD, period doubling; HB, Hopf bifurcation; IPD, inverse period doubling; LPC, limit point bifurcation of cycles.}\label{figure_9}
\end{figure}
\par Solving Eq.\ (\ref{eq.12}), we derive the interior equilibrium point with one real and two complex conjugate eigenvalues. A bifurcation diagram of this interior equilibrium point is plotted by varying $\epsilon$ in the range $[0, 1]$ as shown in Fig.\ \ref{figure_9}(a). The interior equilibrium point is stable until  $\epsilon\approx0.0952$ when it transits to a stable limit cycle via Hopf bifurcation (HB). With increasing $\epsilon \ge 0.0952$, the system evolves into a chaotic state via a period-doubling (PD) cascade. The chaotic state undergoes an inverse period-doubling (IPD) and returns to  a stable steady state at $\epsilon\approx0.8776$ via inverse HB. The evolution of the steady state with $\epsilon$ is verified by plotting the real parts of three eigenvalues of the Jacobian matrix of the system at interior equilibrium point is shown in Fig.\ \ref{figure_9}(b). It is noticed that, for $0\le\epsilon\le0.0952$, real parts of the three eigenvalues $\lambda_{i=1,2,3}$ remain negative which signify stability  of the interior equilibrium  point in this parameter interval. Above   $\epsilon=0.0952$, the real part of the complex conjugate eigenvalues becomes positive  and sustains it up to $\epsilon=0.8776$. In this range of $\epsilon$, the system becomes oscillatory either in periodic [green dots] or in chaotic [blue dots] state [cf. Fig.\ \ref{figure_9}(a)]. At $\epsilon=0.8776$,  the real part of the complex conjugate eigenvalues becomes negative and the equilibrium point turns out to be stable again. So in the interval of  $\epsilon\in(0.0952,0.8776)$, the interior equilibrium point is a saddle focus with a negative real eigenvalue and complex conjugate eigenvalues with positive real parts. The complex dynamics of the system evolves around this equilibrium point and changes with respect to $\epsilon$ values, which is our main focus of study and described in the Sec.\ \ref{origin}, in detail.

\begin{figure}[ht]
\centerline{\includegraphics[scale=0.37]{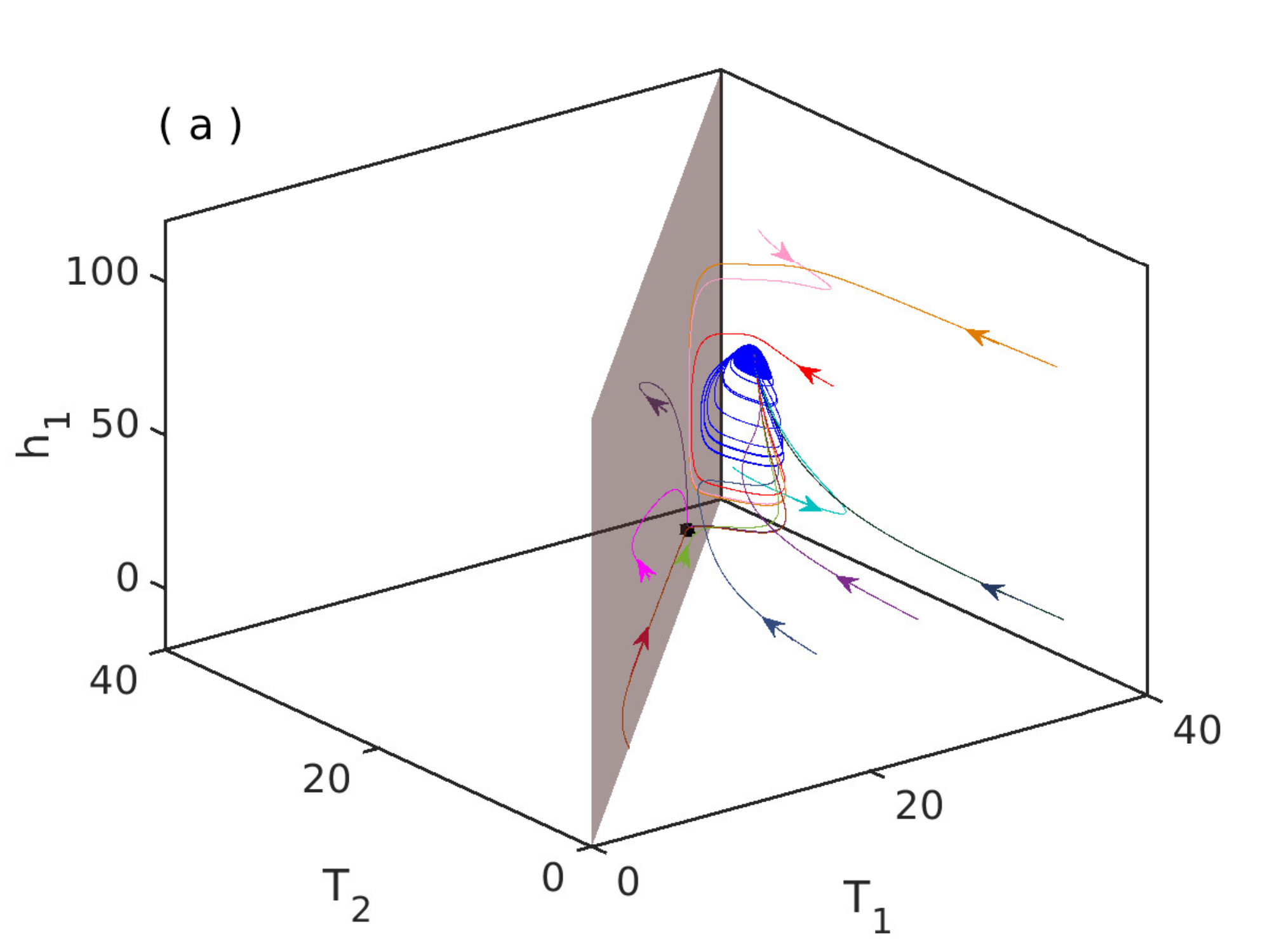}}
\centerline{\includegraphics[scale=0.37]{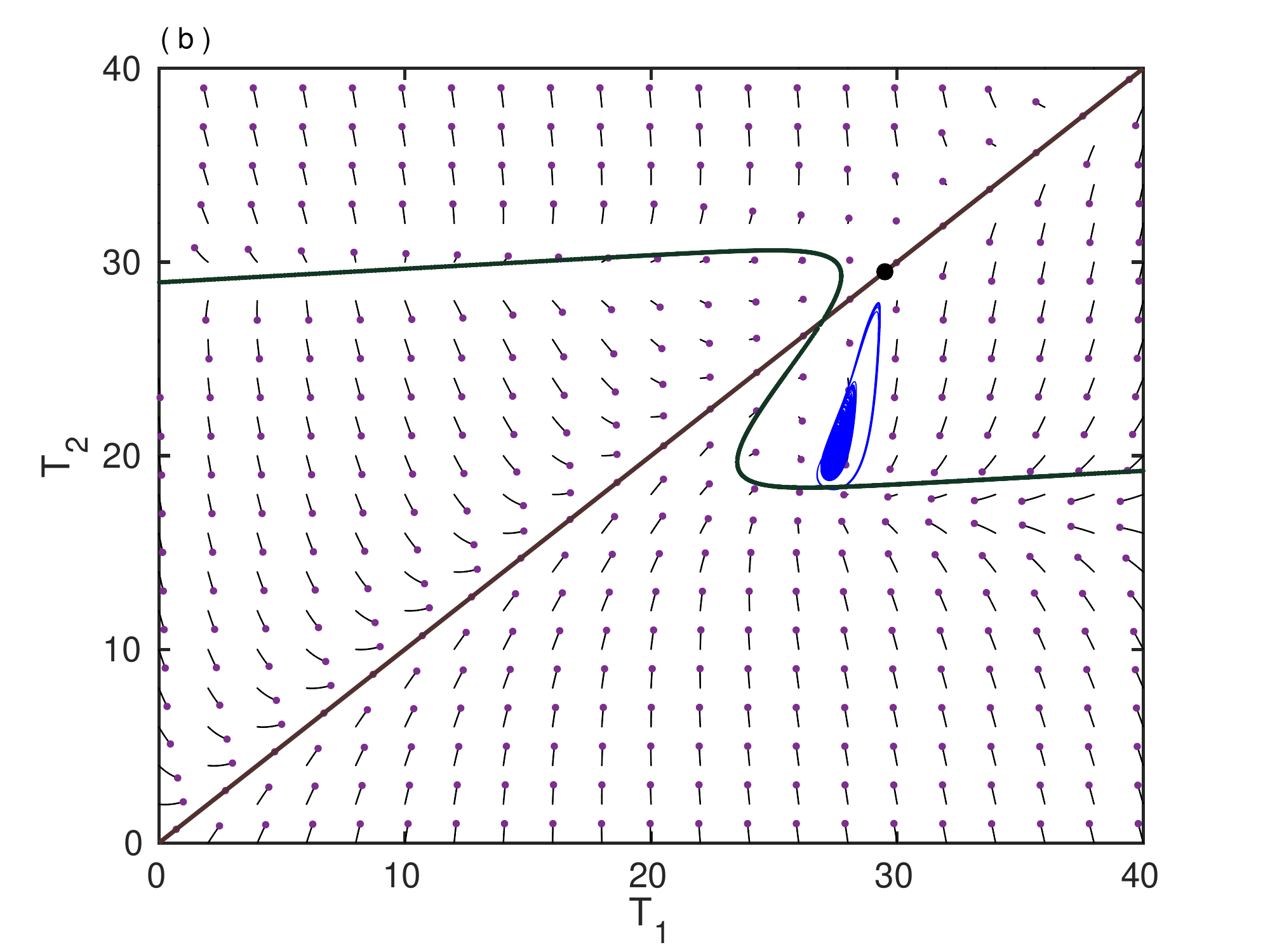}}
\caption{(a) Trajectories of the chaotic attractor (blue line) in 3D plane. Trajectories for different initial conditions are represented by different colors. The diagonal $T_1=T_2$ plane (brown) denotes a stable manifold of the axial equilibrium point $(0,29.5,29.5)$ (black circle). (b) Two-dimensional vector field at $h_1=0$ plane where black line (in Z shape) represents the projection of the critical manifold on this plane. Parameter value: $\epsilon=0.0985$.}\label{figure_10}
\end{figure}
\par The stable manifold, $T_1=T_2$ plane (gray color) is drawn in Fig.\ \ref{figure_10}(a) on which the saddle point $(0,29.5,29.5)$ (black circle) lies. For all initial conditions from the left side of this plane, any trajectory of the system goes unbounded along the unstable manifold of the saddle point $(0,29.5,29.5)$. For  initial conditions from the right-side of this plane, the trajectories of the system converge to the chaotic attractor that is originated around the saddle focus $(h_{1}^*, T_{1}^*, T_{2}^*)$, which lies not far away from the $T_1=T_2$ plane. Some exemplary trajectories are shown in different colors for various choice of initial conditions. The trajectories move toward the attractor along the stable manifold (eigendirection corresponding to the negative real eigenvalue) of the saddle focus, but spirally moves away as repelled by the unstable manifold (corresponding to the complex conjugate eigenvalues with positive real parts) of the saddle focus. This is an ideal situation for the origin of homoclinic chaos, when the trajectory travels a close vicinity of the saddle focus along the stable eigendirection, but  strongly repelled and spiral out along the unstable plane of the saddle focus and makes  large excursions on the same path repeatedly and return to another close vicinity of the saddle focus. Instead, the trajectory of ENSO system is locally unstable in the close vicinity of the saddle focus as usual, but becomes globally unstable too. As a result, the global trajectory is not stable, but traces different paths as shown in Fig.\ \ref{figure_10} (a) that  creates variation in the amplitude of large spiking events and the interspike intervals, which we explain in the Sec.\ \ref{origin}.  The right side of the plane $T_1=T_2$ is the basin of attraction of those events. Fig.\ \ref{figure_10}(b) is a 2D projection of the vector field, which clearly enunciates that all the initial conditions  from the lower-half of the  $T_1=T_2$ line have possibility to originate extreme events. For other initial conditions from the upper-half, i.e., $T_2>T_1$, the system goes unbound along the unstable manifold of the saddle point. In reality, $T_2>T_1$ is also impossible situation. In this way, the equilibrium point $(0,T_r,T_r)$ also plays a significant role to determine the basin of attraction of the extreme events.

\section*{APPENDIX B} 
Here, we have described three PDF. PDF of the Weibull distribution is
\begin{equation}
\begin{array}{lcl}\label{eq.13}
P(r)=\dfrac{k}{\theta}~\Big(\dfrac{r}{\theta}\Big)^{k-1}~e^{-(\frac{r}{\theta})^k},\hspace{15pt}r\in[0,\infty).
\end{array}
\end{equation}
PDF of the Gamma distribution is
\begin{equation}
\begin{array}{lcl}\label{eq.14}
P(r)=\dfrac{1}{\Gamma(k)\theta^k}~r^{k-1}e^{-\frac{r}{\theta}},\hspace{15pt}r\in(0,\infty).
\end{array}
\end{equation}
PDF of the Log-normal distribution is
\begin{equation}
\begin{array}{lcl}\label{eq.15}
P(r)=\dfrac{1}{\sqrt{2\pi}r\sigma}~exp(\dfrac{-(\log r-\mu)^2}{2\sigma^2}), \hspace{15pt}r\in(0,\infty).
\end{array}
\end{equation}
Here, $k(>0)$ and $\theta(>0)$ are shape and scale parameters for the first two distributions [\ref{eq.13} and \ref{eq.14}], $\mu$ and $\sigma(>0)$ are mean and standard deviation for the last one \eqref{eq.15}. In this context, $r$ represents interevent interval. We estimate the parameters which we have recorded are given in the Table I: 

\begin{center}
	Table I
	\begin{tabular}{ |c|c|c|c|c| } 
		\hline
		 & $\epsilon=0.0985$ & $\epsilon=0.0988$ \\
		\hline
		\multirow{2}{10em}{Weibull distributon} & $\theta$=23463.1 & $\theta$=15815.20 \\ 
		& $k$=1.49827 & $k$=3.5059 \\ 
		\hline
			\multirow{2}{10em}{Gamma distributon} & $\theta$=7322.98 & $\theta$=706.46 \\ 
		& $k$=2.8527 & $k$=20.4212 \\ 
		\hline
				\multirow{2}{10em}{Log-normal distributon} & $\mu=9.7617$ & $\mu=9.5522$ \\ 
		& $\sigma=0.5487$ & $\sigma=0.2122$ \\ 
			\hline
	\end{tabular}
\end{center}

\section*{Appendix C: Box plot analysis}\label{appendix_c}
\begin{figure}[ht]
	\centerline{\includegraphics[scale=0.4]{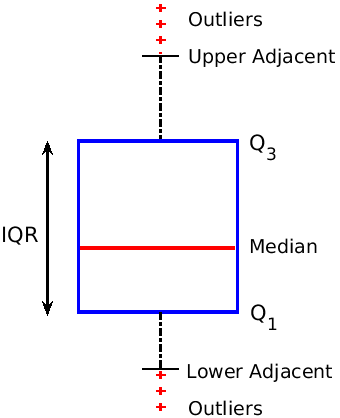}}
	\caption{ Schematic diagram of a box plot: A blue box is drawn from the first quartile (Q$_1$) to the third quartile (Q$_3$). Red horizontal straight line is traced at the median of the data set. 
Two black dashed vertical lines are drawn respectively from {\it lower adjacent} to Q$_1$ and Q$_3$ to {\it upper adjacent}. Red (+) markers denote the outlier events, which are least predictable. 
}
   \label{figure_7_a}
\end{figure}
\par A box plot \cite{box} is a standard contrivance to display the distribution of a given data. It basically produces a five measures summary, namely {\it lower adjacent}, first quartile $Q_1$, second quartile ({\it median}), third quartile $Q_3$, and {\it upper  adjacent}. A box plot gives more information than the measures of central tendency. For information on variability or dispersion of IEI, it gives an  impression about the spread of IEI values. Here {\it median} is the second quartile, which is the middle-most value of the dataset. The first quartile Q$_1$, is the middle point between the {\it median} and the smallest number of the whole data set.  While the third quartile Q$_3$ is the middle point between the {\it median} and the highest value of the whole data set. IQR is defined as the difference of Q$_3$ and Q$_1$. The {\it lower  adjacent} and {\it upper  adjacent} are not always the smallest and largest values in the data set, respectively. Here, {\it lower  adjacent} is the smallest value of the data set or ($\mbox{Q}_1-1.5\times\mbox{IQR}$), whichever is the maximum. While {\it upper adjacent} is the minimum of the maximum value of the data set and ($\mbox{Q}_3+1.5\times\mbox{IQR}$). The events from {\it lower  adjacent} to {\it upper adjacent} are predictable events, and the range of {\it upper adjacent} to {\it lower adjacent} is known as the range of predictable events. Although, we use large size of data-set as available from our simulations and as necessary for such kind of box-analysis. 

\par A schematic diagram of the box plot is shown in Fig. \ref{figure_7_a}. 
The five measures approximately divide the entire data set into four sections, each one approximately contains $25\%$ of the data sets. Beyond the {\it upper adjacent} and {\it lower adjacent} values, events are outliers, which are seemingly unpredictable and much occurred rarely. The outliers (red plus) tell us which events are out of our predictable range. The probability of occurrence of an event within the interval [{\it lower adjacent}, Q$_1$] is $0.25$. While the probability of an event that lies within the intervals [{\it lower adjacent}, median] and [{\it lower adjacent}, Q$_3$] are $0.5$ and $0.75$, respectively.   It is almost sure to lie the predictable events within the interval [{\it lower adjacent}, {\it upper adjacent}]. Five measures of box plot corresponding to the Fig.\ \ref{figure_7_b}(b) and Figs.\ \ref{figure_7_c}(a) and \ref{figure_7_c}(d) are given in the Table II:  
\begin{widetext}
	\begin{center}
		Table II
		\begin{tabular}{ |c|c|c|c|c|c|c| }
			\hline
			Box plot & Lower Adjacent & $Q_{1}$ & Median & $Q_{3}$ & Upper Adjacent & Outliers \\
			\hline
			\multirow{1}{5em}{Fig.~\ref{figure_7_b}(b)} & 12604.38 & 13291.14 & 22822.3 & 46548.3 & 96434.04 & 6.55\% \\
			\hline
			\multirow{2}{5em}{Fig.~\ref{figure_7_c}(a)} & 12604.38 & 12615.708 & 12626.15 & 12641.57 & 12659.5 & 0.0 \\
			& 13267.38 & 13276.74 & 13284.7 & 13299.73 & 13322.18 & 0.0 \\
			\hline
			\multirow{1}{5em}{Fig.~\ref{figure_7_c}(b)} & 14980.68 & 23718.93 & 36832.34 & 62234.63 & 120008.18 & 5.05\% \\
			\hline
			\multirow{1}{5em}{Fig.~\ref{figure_7_c}(c)} & 8345.9 & 11303.59 & 12609.16 & 13275.38 & 13322.18 & 5.74\% \\
			\hline
			\multirow{1}{5em}{Fig.~\ref{figure_7_c}(d)} & 8955.99 & 16956.65 & 22821.38 & 37095.83 & 67282.32 & 5.64\% \\
			\hline
			\end{tabular}
		\end{center}
\end{widetext}

\end{document}